%% file: anyons_qw.tex
\documentclass[%
 reprint,
 amsmath,amssymb,
 aps,
 longbibliography,
floatfix,
]{revtex4-2}

\usepackage{graphicx}
\usepackage{dcolumn}
\usepackage{bm}

\PassOptionsToPackage{hyphens}{url}
\usepackage[hidelinks]{hyperref}
\hypersetup{breaklinks=true}

\begin{document}

\title{Realization of 1D Anyons with Arbitrary Statistical Phase}

\author{Joyce~Kwan$^{1}$}
\author{Perrin~Segura$^{1}$}
\author{Yanfei~Li$^{1}$}
\author{Sooshin~Kim$^{1}$}
\author{Alexey~V.~Gorshkov$^{2}$}
\author{Andr\'e~Eckardt$^{3}$}
\author{Brice~Bakkali-Hassani$^{1}$}
\author{Markus~Greiner$^{1}$}
    \email{mgreiner@g.harvard.edu}
\affiliation{
$^{1}$Department of Physics, Harvard University, Cambridge, MA 02138, USA \\
$^{2}$Joint Quantum Institute and Joint Center for Quantum Information and Computer Science, NIST/University of Maryland, College Park, MD 20742, USA \\
$^{3}$Institut f\"ur Theoretische Physik, Technische Universit\"at Berlin, Berlin 10623, Germany
}

\date{\today}

\begin{abstract}
Low-dimensional quantum systems can host anyons, particles with exchange statistics that are neither bosonic nor fermionic. Despite indications of a wealth of exotic phenomena, the physics of anyons in one dimension (1D) remains largely unexplored. Here, we realize Abelian anyons in 1D with arbitrary exchange statistics using ultracold atoms in an optical lattice, where we engineer the statistical phase via a density-dependent Peierls phase. We explore the dynamical behavior of two anyons undergoing quantum walks, and observe the anyonic Hanbury Brown-Twiss effect, as well as the formation of bound states without on-site interactions. Once interactions are introduced, we observe spatially asymmetric transport in contrast to the symmetric dynamics of bosons and fermions. Our work forms the foundation for exploring the many-body behavior of 1D anyons.
\end{abstract}

\maketitle

\include{main_text}

\include{supplementary}

\end{document}

%% file: main_text.tex
\section{Introduction}
\begin{figure*}
\includegraphics{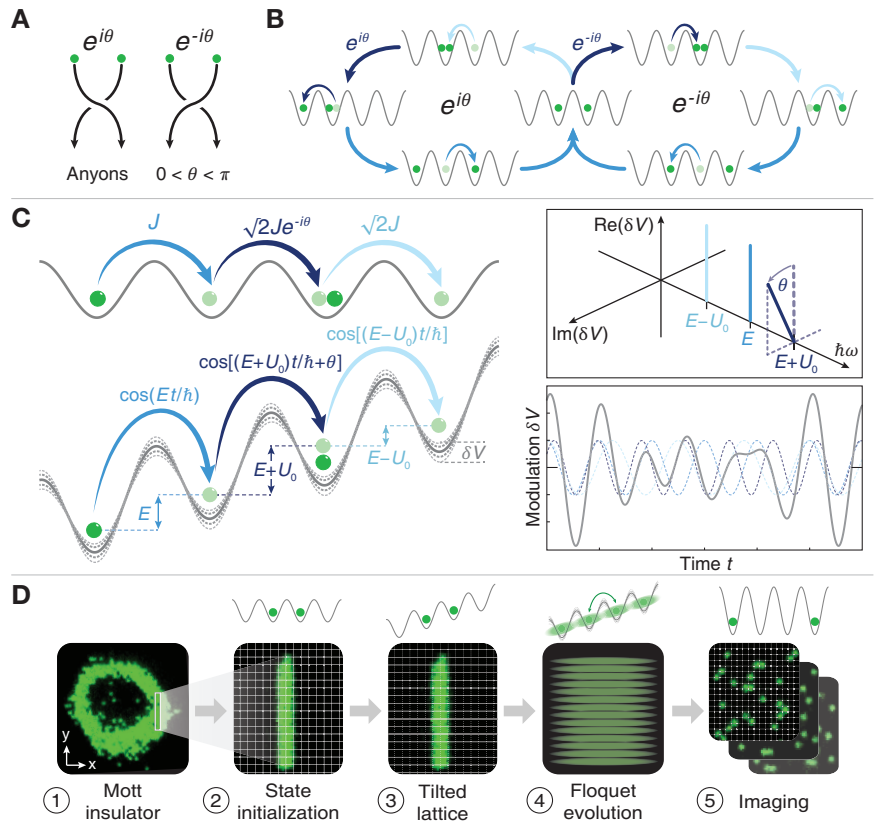} 
\caption{\textbf{Realization of anyons in 1D.} (A) Abelian anyons have an exchange phase that interpolates between $0$ (bosons) and $\pi$ (fermions). (B) In 1D, the wavefunction acquires phase $-\theta$ ($\theta$) when a particle tunnels right (left) through an occupied site, analogous to clockwise (counter-clockwise) exchange in 2D. (C) We realize the anyon-Hubbard model (AHM) in a tilted optical lattice with energy offset $E$ per site to suppress tunneling, then induce tunneling by modulating the lattice depth with three frequency (3-tone) components, each with amplitude $\delta V$: $E$ to tunnel from a singly-occupied to an empty site, $E+U_0$ to tunnel from a singly-occupied to a singly-occupied site, and $E-U_0$ to tunnel from a doubly-occupied to an empty site, where $U_0$ is on-site interaction in the initial Hamiltonian \cite{anyons_sm}. Offsetting the phase of component $E+U_0$ by $\theta$ realizes the density-dependent Peierls phase. Insets: 3-tone modulation in frequency (top) and time (bottom), the grey line is the sum of the three components. (D) Experimental sequence: (1-2) initialize two columns of atoms from a Mott insulator of \textsuperscript{87}Rb; (3) tilt the lattice, then lower its depth $V_x$ to prepare for modulation-induced tunneling; (4) abruptly apply 3-tone modulation to induce several independent quantum walks along $x$; (5) project to the number basis and perform fluorescence imaging \cite{anyons_sm}.}
\label{fig:protocol}
\end{figure*}

In three dimensions, quantum theory admits two types of particles, bosons and fermions, depending on whether the many-body wavefunction acquires a phase $\theta$ of $0$ (bosons) or $\pi$ (fermions) when two indistinguishable particles exchange positions. In practice, this means bosons prefer to occupy the same quantum state, such as photons in a laser or atoms forming a Bose-Einstein condensate, while fermions obey the Pauli exclusion principle, such as electrons occupying different orbitals to produce elements in the periodic table. When dimensions are reduced to two or lower, the exchange phase $\theta$ can interpolate between the bosonic and fermionic limits, leading to fractional statistics (Fig.\,\ref{fig:protocol}A) \cite{leinaas_theory_1977,Wilczek_1982,Greiter_2022}. Particles with such an exchange phase are called anyons because they can acquire \textit{any} phase \cite{Wilczek_1982}. 

Typically known in two dimensions (2D), anyons have gained immense interest in the contexts of fractional quantum Hall states, where they arise as quasiparticle excitations \cite{Halperin_1984,Arovas_1984,Bartolomei_2020, Nakamura_2020}, and fault-tolerant quantum computation, where they serve as a key building block \cite{Kitaev_2003,Bravyi_2006,Nayak_2008,Clarke_2013,google_non_abelian_2022}. In one dimension (1D), the existence of anyons was established when theoretical and experimental work showed spinon excitations in a Heisenberg antiferromagnetic chain obey a fractional exclusion principle \cite{Haldane_1991,Greiter_2009,Mourigal_2013}. Models of 1D systems with fractional statistics have been proposed in the continuum \cite{1999_Kundu,Harshman_2020,2021_Bonkhoff} and on a discrete lattice \cite{Keilmann_2011}. In this work, we focus on the lattice model, which indicates a wealth of exotic phenomena, including asymmetric momentum distributions \cite{Hao_2008,Hao_2009,Tang_2015}, the continuous buildup of Friedel oscillations with increasing $\theta$ \cite{2016_Strater,Yuan_2017}, a Mott insulator to superfluid phase transition induced by statistical parameter $\theta$ \cite{Keilmann_2011}, and a novel two-component superfluid phase \cite{2015_Greschner,2017_Zhang}. Despite these intriguing prospects, the physics of 1D anyons remains largely unexplored in experiment \cite{Sansoni_2012,Matthews_2013,zhang_observation_2022}.

Here, we realize Abelian anyons in 1D with arbitrary statistical phase using ultracold \textsuperscript{87}Rb atoms in an optical lattice. We leverage the precision and control of a quantum gas microscope \cite{Bakr_2009} to imprint the statistical phase in a deterministic way and explore dynamical behavior via two-particle quantum walks \cite{Wang_2014,Preiss_2015}. The system is governed by the anyon-Hubbard model (AHM) \cite{Keilmann_2011}, which we realize by engineering an equivalent model, the Bose-Hubbard model (BHM) with density-dependent phase \cite{Clark_2018,G_rg_2019,Schweizer_2019}, 
\begin{equation} \label{eq:hamiltonian}
{\mathcal{H}} = -J\sum_j \left({b}_j^\dagger e^{-i {n}_j \theta} {b}_{j-1} + \textrm{h.c.} \right) + \frac{U}{2} \sum_j {n}_j ({n}_j - 1),
\end{equation}
where ${b}_j^\dagger$ (${b}_j$) is the bosonic creation (annihilation) operator, ${n}_j = {b}_j^\dagger {b}_j$ is the particle number operator, $J$ is the tunneling amplitude between neighboring sites, $U$ is the on-site, pairwise repulsive interaction energy, and $-{n}_j \theta$ is the density-dependent Peierls phase acquired upon tunneling right from site $j-1$ to site $j$. Note that the density-dependent phase acquired upon tunneling left, encapsulated in the Hermitian conjugate, is ${n}_j \theta$; hence this model breaks spatial inversion symmetry, a property associated with fractional statistics \cite{Wilczek_1990}. The BHM with density-dependent phase can be mapped to the AHM by a generalized Jordan-Wigner transformation \cite{Keilmann_2011,anyons_sm}.

We can gain intuition for anyons in 1D lattices by recognizing similarities with their 2D counterparts. Anyons in 1D can be considered as bosons that create a gauge potential in the form of the density-dependent Peierls phase for other particles, analogous to the charge-flux tube composites that exemplify anyons in the 2D theory \cite{Greiter_2022}. As a result, two particles can traverse through states forming a closed loop in Fock space and acquire a phase $\theta$ corresponding to a geometric phase (Fig.\,\ref{fig:protocol}B). This process offers an analogy to braiding in 2D, where tunneling right (left) through an occupied site, which exchanges positions between particles and aquires phase $-\theta$ ($\theta$), corresponds to clockwise (counter-clockwise) exchange. 

\begin{figure*}
\includegraphics{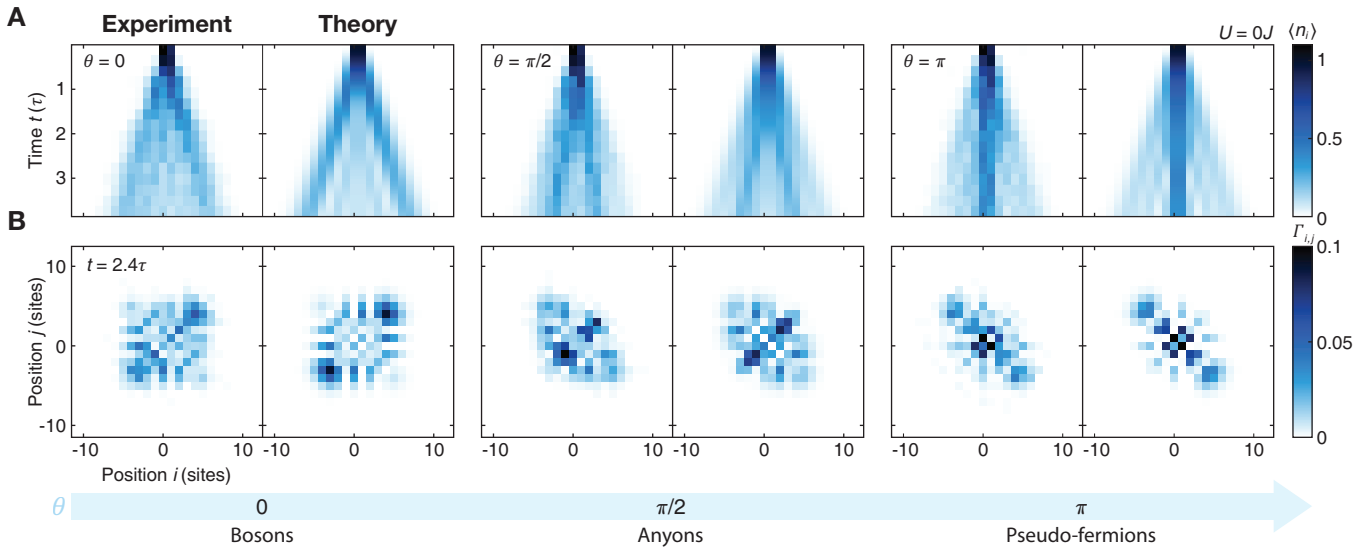}
\caption{\textbf{Quantum walks of two anyons, $U=0$.} (A) Density profile of two-particle quantum walks for various $\theta$, each obtained by averaging over $\sim 1800$ experimental runs. Good agreement with theory shows coherence up to our experiment time $t\approx 4\tau$ (in units of inverse tunneling time $\tau = 15.0(3)$ ms) across $\sim 20$ sites. (B) Density-density correlator $\Gamma_{i,j}$ at $t=2.40(5)\tau$. When $\theta=0$, bosonic bunching appears as weights along the diagonal $i=j$. When $\theta=\pi$, anti-bunching behavior of pseudo-fermions appears as weights along the anti-diagonal $i=-j$. When $\theta=\pi/2$, $\Gamma_{i,j}$ reveals fractional statistics, showing both strong diagonal weights and the onset of fermionization.}
\label{fig:quantum_walk}
\end{figure*}

\section{Experimental protocol}
We realize the BHM with density-dependent phase via Floquet engineering by modulating a tilted lattice with three frequency (3-tone) components to induce occupation-dependent tunneling processes (Fig.\,\ref{fig:protocol}C) \cite{2016_Cardarelli}. Specifically, a magnetic field gradient produces an energy offset $E$ between lattice sites to suppress tunneling, then tunneling is reintroduced by modulating the lattice depth with three frequencies, each with amplitude $\delta V$: (1) $E$ to tunnel from a singly-occupied site to an empty site, (2) $E+U_0$ to tunnel from a singly-occupied site to a singly-occupied site, and (3) $E-U_0$ to tunnel from a doubly-occupied site to an empty site, where $U_0$ is the interaction energy in the initial Hamiltonian \cite{anyons_sm}. The amplitude of modulation $\delta V$ determines $J$, and offsetting the phase of frequency component $E+U_0$ by $\theta$ from components $E$ and $E-U_0$ realizes the density-dependent phase and therefore the statistical parameter. Modulating the lattice with these frequencies, which are resonant with parameters of the initial Hamiltonian, realizes the BHM with density-dependent phase corresponding to the non-interacting AHM. We can engineer an effective on-site interaction $U$ in the AHM by detuning the sidebands to become $E-(U_0-U)$ and $E+(U_0-U)$ \cite{2016_Cardarelli}. 

For the experiments that follow, we employ the single-site control of our quantum gas microscope to study the dynamics of two anyons undergoing quantum walks, with and without interaction $U$ in the AHM. Using a digital micromirror device \cite{Zupancic_2016}, we initialize two columns of atoms along $y$ in a deep optical lattice with $V_x = V_y = 45 E_\textrm{R}$, where $E_\textrm{R} = h \times 1.24$ kHz for our lattice constant of $a=680$ nm and $h = 2\pi\hbar$ is Planck's constant, in preparation for several independent quantum walks of two particles along $x$ (Fig.\,\ref{fig:protocol}D). We ramp a magnetic field gradient to offset lattice sites by $E$, then lower $V_x$ to $4 E_\textrm{R}$ and abruptly turn on 3-tone modulation with $\delta V = 20\% \times V_x$ to induce quantum walks along $x$. The parameters for the experiments are $J/h = 10.6(2)$ Hz, $U_0/h = 210(4)$ Hz, and $E/h = 800(3)$ Hz, and we measure time in units of inverse tunneling rate $\tau = h/(2\pi J) = 15.0(3)$ ms. We detect doubly-occupied sites by ramping the magnetic field gradient past $U_0$ to separate atoms before imaging for half of each data set, circumventing pairwise loss of atoms due to light-assisted collisions \cite{anyons_sm}. 

\section{Experimental results} 
We first study anyonic behavior by measuring the quantum correlations of two particles simultaneously undergoing quantum walks when $U=0$. The quantum walk of two particles is sensitive to quantum statistics due to the Hanbury Brown-Twiss effect, where all two-particle processes add coherently to develop quantum correlations \cite{Henny_1999,Jeltes_2007,Peruzzo_2010,Sansoni_2012,Preiss_2015}. Initializing the state as ${b}_0^\dagger {b}_1^\dagger |0\rangle = |...0110...\rangle$, we capture the trajectory by evolving the system for successively longer periods of time before imaging and average over many images to obtain the probability distribution of the two particles (Fig.\,\ref{fig:quantum_walk}A). We capture the quantum walk trajectory for various $\theta$ and characterize quantum statistics using the density-density correlator $\Gamma_{i,j} = \langle{b}_j^\dagger {b}_i^\dagger {b}_i {b}_j\rangle$ (Fig.\,\ref{fig:quantum_walk}B). When $\theta=0$, bosonic bunching appears as weights along or near the diagonal $i=j$ of the correlation matrix $\Gamma_{i,j}$  \cite{Preiss_2015}, consistent with Bose-Einstein statistics. When $\theta=\pi$, weights appear along or near the anti-diagonal $i=-j$, indicating anti-bunching behavior emblematic of fermions. Weights appear along the diagonal $i=j$ because bosons now behave as pseudo-fermions, which do not obey the Pauli principle on-site, acting as fermions off-site and bosons on-site \cite{Keilmann_2011}. Therefore, their spatial correlations differ from those of true fermions, but nonetheless show the essential trait of anti-bunching. Finally, for fractional phase $\theta=\pi/2$, the correlation matrix $\Gamma_{i,j}$ shows intermediate levels of bunching and anti-bunching to reveal fractional statistics; strong weights appear along the diagonal while off-diagonal weights indicate the onset of fermionization. Good agreement with theory shows the system maintains coherence up to our experiment time of $t \approx 4\tau$ across $\sim 20$ sites. Unless otherwise noted, all theoretical predictions were obtained ab initio using exact diagonalization of the BHM with density-dependent phase, with Hubbard parameters determined from the calibrated value of effective tunneling $J$.

\begin{figure}[t]
\includegraphics{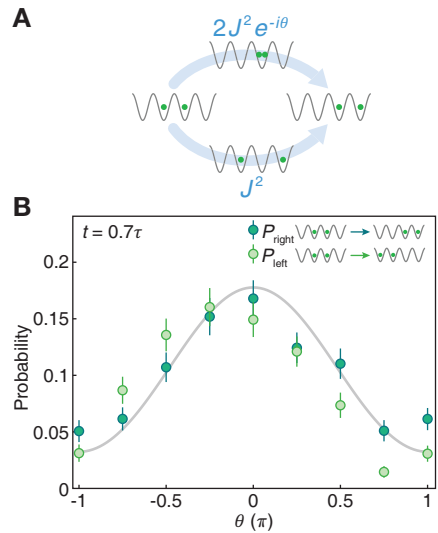}
\caption{\textbf{Interferometric picture of Fock state evolution.} (A) We can understand the effect of the statistical phase $\theta$ on tunneling processes by appealing to an interferometric interpretation of Fock state evolution. The initial state ($|...0110...\rangle$) splits into two arms, acquiring phase $-\theta$ in the upper arm ($|...0020...\rangle$) before interfering with the lower arm ($|...0101...\rangle$) to arrive at the final state ($|...0011...\rangle$). (B) Probability $P_{\textrm{right}}$ of occupation of $|...0011...\rangle$, corresponding to both atoms having tunneled one site right, after $t=0.70(2)\tau$ as a function of $\theta$. The same relation holds for $P_{\textrm{left}}$, the probability for both atoms to tunnel one site left after $t=0.70(2)\tau$. Decrease in probability as $\theta$ approaches $\pm\pi$ can be understood as development of destructive interference between paths in Fock space, maximally cancelling when $\theta=\pm\pi$ to localize atoms on their initial sites. Solid line shows prediction from theory. Errorbars denote the s.e.m.}
\label{fig:interferometer}
\end{figure}

A bimodal structure emerges in the density profiles, with an internal cone that narrows as $\theta$ increases from $0$ to $\pi$ amidst a background density. We can understand this structure by appealing to an interferometric interpretation of Fock state evolution (Fig.\,\ref{fig:interferometer}A). Our initial state ($|...0110...\rangle$), the source, splits into upper ($|...0020...\rangle$) and lower ($|...0101...\rangle$) arms to interfere at the final state ($|...0011...\rangle$), a process that corresponds to both atoms tunneling one site right after short time evolution $t<\tau$. When $\theta=0$, the two arms constructively interfere to arrive at the final state with enhanced probability, but when $\theta=\pi$, the two arms destructively interfere, reducing the path to the final state, and the system is more likely to remain in the initial state. The same picture applies for tunneling leftward, hence as $\theta$ increases from $0$ to $\pi$, atoms are less likely to delocalize, forming the strong density pattern in the center in the pseudo-fermion limit. Importantly, because the states comprising the interferometer are the same as those in the loop in Fock space describing particle exchange (Fig.\,\ref{fig:protocol}B), interference between the two arms directly reflects anyonic exchange statistics. We measure the Fock state distribution at short time $t=0.70(2)\tau$ and observe that the proportion $P_{\textrm{right}}$ of experimental runs in state $|...0011...\rangle$ decreases as $\theta$ changes from $0$ to $\pm\pi$, as explained by the development of destructive interference in the interferometric picture of Fock state evolution (Fig.\,\ref{fig:interferometer}B). Since the same interferometric picture applies for tunneling leftward, the proportion $P_{\textrm{left}}$ of experimental runs in $|...1100...\rangle$ at $t < \tau$ is approximately equal to $P_{\textrm{right}}$, also decreasing as $\theta$ changes from $0$ to $\pm\pi$ due to destructive interference. 

\begin{figure*}
\includegraphics{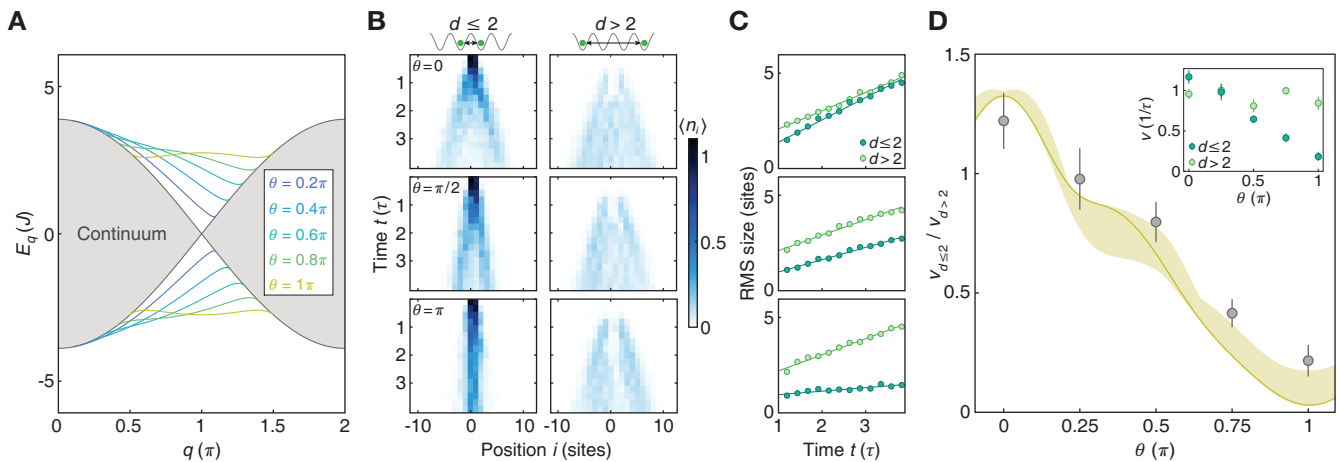}
\caption{\textbf{Characterizing bound pairs of anyons, $U = 0$.} (A) Two-particle spectrum $E_q$, where $q$ is the center-of-mass quasimomentum, with scattering states forming a continuum (grey-shaded region) and bound states detaching from the continuum (solid lines, from blue to yellow: $\theta = m \pi / 5$, $1 \le m \le 5$). (B) Density distributions conditioned on the relative distance $d$ between the two particles undergoing a quantum walk (same data as in Fig.\,\ref{fig:quantum_walk}): left (right) plots correspond to near (distant) particles separated by $d\le 2$ ($d > 2$) sites, with $\theta = 0, \pi/2, \pi$, from top to bottom. (C) Root-mean-squaure (RMS) size evolution of the near (dark green) and distant (light green) particle components, with $\theta = 0, \, \pi/2, \, \pi$ from top to bottom. The spreading velocities $v_{d \le 2}, v_{d > 2}$ are defined as the slopes of the linear fits (solid lines). Error bars (smaller than data points) were obtained from a bootstrap analysis. (D) Ratio of spreading velocities $v_{d \le 2} / v_{d > 2}$ as a function of $\theta$. Solid line shows theoretical prediction, with shaded region corresponding to uncertainty in tilt calibration \cite{anyons_sm}. Inset: Spreading velocities $v_{d \le 2}$, $v_{d > 2}$ as a function of $\theta$. Errorbars denote the s.e.m.}
\label{fig:bound_state}
\end{figure*}

The narrowing internal cone in the density profiles indicates the formation of bound states as $\theta$ increases from $0$ to $\pi$ even in the absence of on-site interaction $U$ \cite{2006_Winkler,2013_Fukuhara,2022_Kranzl}. This phenomenon occurs because the density-dependent gauge field mediates interactions between bosons \cite{2016_Cardarelli,2017_Zhang,2018_Greschner}. Theoretically, the two-particle spectrum of the AHM with $U = 0$ consists of a continuum of scattering states surrounded by two branches of bound states for $\theta \neq 0$, an upper branch with energy $E_q > 0$ and lower branch with $E_q < 0$, where $q$ is the center-of-mass quasimomentum (Fig.\,\ref{fig:bound_state}A). Each branch shows a preferred direction of propagation, given by the sign of the group velocity $\textrm{d}E_q/\textrm{d}q$. Our initial state $|...0110...\rangle$ projects onto both branches of bound states with equal weight and onto scattering states; therefore, the internal cone in the density profiles appears symmetric about the center \cite{anyons_sm}. 

We distinguish the formation of bound pairs from scattering states by analyzing the spreading velocities of the two wavefunction components. First, we characterize the two distinct dynamics by conditioning the density profiles in Fig.\,\ref{fig:quantum_walk} on the relative distance $d$ between the particles, analyzing separately the spatial distributions of near ($d \le 2$ sites) and distant ($d > 2$ sites) particles (Fig.\,\ref{fig:bound_state}B); for analyses conditioned on different relative distances, which show similar behavior, see \cite{anyons_sm}. Then, we determine the root mean square (RMS) size of each component as a function of time (Fig.\,\ref{fig:bound_state}C) and perform a linear fit to extract the spreading velocities $v_{d \le 2}$ and $v_{d > 2}$ (Fig.\,\ref{fig:bound_state}D, inset). We see $v_{d > 2}$ is approximately independent of $\theta$ as expected for scattering states, whereas $v_{d \le 2}$ strongly decreases as $\theta$ increases from $0$ to $\pi$. This behavior is consistent with the narrowing internal cone in the density profiles and the decreasing group velocity $\textrm{d}E_q/\textrm{d}q$ in the two-particle spectrum as $\theta$ increases from $0$ to $\pi$. Note that we systematically extract a slightly reduced velocity compared to theory due to error in calibrating site offset $E$, which results in a residual tilt in the effective model \cite{anyons_sm}. Therefore, we characterize the formation of bound pairs with the ratio $v_{d \le 2} / v_{d > 2}$, a quantity more robust to a residual tilt (Fig.\,\ref{fig:bound_state}D). Data points at $\theta=\pi/4, \, 3\pi/4$ correspond to the analysis of density profiles of two-particle quantum walks subject to these phases \cite{anyons_sm}. Good agreement with theory further demonstrates our measurements show the existence of bound pairs in the absence of on-site interactions. 

\begin{figure*}
\includegraphics{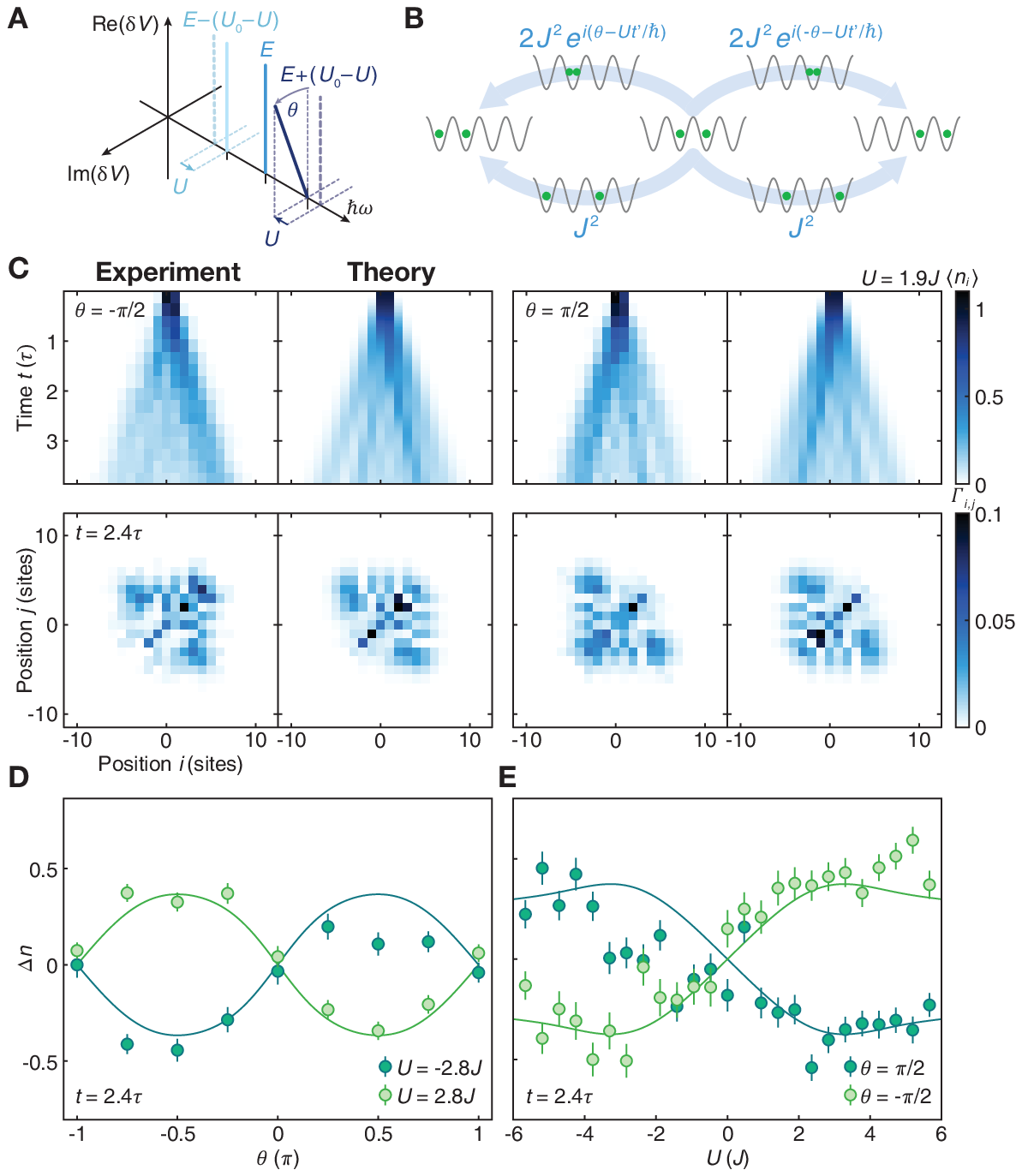}
\caption{\textbf{Asymmetric transport due to the presence of on-site interaction $U$.} (A) Introducing repulsive on-site $U$ in the AHM amounts to detuning sidebands $E-U_0$ and $E+U_0$ by $U$ toward the center frequency $E$. (B) Presence of $U$ breaks inversion symmetry in the density profiles of quantum walks for fractional $\theta$. Now $P_{\textrm{right}} \neq P_{\textrm{left}}$ because the phase accumulated by tunneling right through an occupied site differs from that accumulated by tunneling left through an occupied site. (C) For $U=1.9(4)J$ and $\theta=-\pi/2$, the density profile shows rightward trajectory and $\Gamma_{i,j}$ at $t = 2.40(5)\tau$ shows strong spatial correlations with the right half of the quantum walk. Transport changes direction toward the left for $\theta=\pi/2$. (D) For constant $U = \pm 2.8(4) J$, direction of transport as a function $\theta$ summarized by $\Delta n$, difference in atom number between right and left halves of the quantum walk. It changes with the sign of $\theta$, and with the sign of $U$, with $\Delta n \approx 0$ for $\theta=0, \pi$, consistent with symmetric expansion of bosons and pseudo-fermions. (E) For constant $\theta$, direction of transport as a function of $U$. Direction of transport depends on the sign and the strength of $U$. Errorbars denote the s.e.m.}
\label{fig:asymmetric_transport}
\end{figure*}

A defining characteristic of 1D anyons is spatially asymmetric transport when interactions are present because the AHM is not inversion symmetric \cite{Liu_2018}. Inversion symmetry is broken because phase $-\theta$ ($\theta$) is acquired when a particle tunnels right (left) through an occupied site, a property that becomes apparent in the density profile of anyonic quantum walks when $U \neq 0$ \cite{Liu_2018}. This can be understood by referring to the interferometric picture of Fock states at short time $t < \tau$ (Fig.\,\ref{fig:asymmetric_transport}B). Phases acquired in the upper arms $|...0200...\rangle$ and $|...0020...\rangle$ are now respectively $\theta - U t'/\hbar$ and $-\theta - U t'/\hbar$, where $t'=t/3$ \cite{anyons_sm}, which for $\theta=0$ or $\pi$, result in equal probability to arrive at final states $|...1100...\rangle$ and $|...0011...\rangle$. When $\theta$ is fractional, the probabilities to arrive at final states $|...1100...\rangle$ and $|...0011...\rangle$ are not equal, with preference to tunnel in a particular direction. This picture also explains when the sign of $\theta$ or $U$ changes, so does the direction of transport.

Introducing an effective on-site interaction in our scheme amounts to detuning the sidebands $E-U_0$ and $E+U_0$ by $U$ (Fig.\,\ref{fig:asymmetric_transport}A), which, when present, leads to asymmetric transport in the density profile of two anyons undergoing quantum walks (Fig.\,\ref{fig:asymmetric_transport}C). The density profile shows transport toward the right when $\theta=-\pi/2$ and $U=1.9(4) J$, with $\Gamma_{i,j}$ at $t = 2.40(5)\tau$ showing correlations with the right half of the quantum walk. For opposite phase $\theta=\pi/2$, the direction of transport is now toward the left, corresponding to a reversal of phases accumulated in the left and right upper arms of the interferometer. We quantify the asymmetry of transport by the difference in atom number between the right and left halves of the quantum walk, $\Delta n = \sum_{i > 0} \langle n_i \rangle - \sum_{i \leq 0} \langle n_i \rangle$, after some time evolution. At constant $U$, we see transport is asymmetric for fractional $\theta$ and changes direction as $\theta$ reverses sign (Fig.\,\ref{fig:asymmetric_transport}D). When $\theta=0$ or $\pi$, we measure $\Delta n \approx 0$, consistent with the symmetric density profiles of bosons and pseudo-fermions. At constant fractional $\theta$, the direction of transport changes with the sign of $U$ (Fig.\,\ref{fig:asymmetric_transport}E), behavior inherent only to anyons, as expansion dynamics of bosons and fermions are identical for $\pm U$ \cite{Schneider_2012,Ronzheimer_2013,Yu_2017}. General agreement with theory shows interactions can be engineered across a broad range, $-6J < U < 6J$, with deviations appearing when $U<0$ due to Floquet heating \cite{anyons_sm}.

\section{Discussion}
In summary, we engineer a density-dependent Peierls phase to realize 1D anyons with tunable exchange phase and reveal fractional statistics in the Hanbury Brown-Twiss effect of two-anyon quantum walks. We show this density-dependent phase, a form of interaction, is the mechanism behind the formation of bound states even in the absence of on-site interactions. Then, once we introduce on-site interactions, the breaking of inversion symmetry, a property associated with fractional statistics, becomes apparent in the density profiles due to the interplay between the density-dependent phase and on-site interactions.

The 3-tone Floquet scheme that realizes the AHM expands existing capabilities of Hamiltonian engineering, enabling control of $U$ without Feshbach resonances and the simulation of a broad class of Hubbard models due to the ability to independently control $J$, $U$, and $\theta$. Floquet engineering with ultracold atoms is generally a challenge, requiring cancellation of coupling to dissipative modes \cite{Viebahn_2021} or an optimal balance between driving parameters and coherence time of the system \cite{Eckardt_2017}. Yet our post-selection rate remains relatively high at $\sim 60\%$ at the end of a typical experiment lasting $t \approx 4\tau$ despite the number of modulation components \cite{anyons_sm}. It would therefore be viable to expand the scheme, such as by increasing the number of modulation components or by dynamically changing effective Hubbard parameters.

The many-body behavior of an ensemble of 1D anyons is a promising direction for future study. For example, the bound states we observe play a crucial role in the emergence of a novel superfluid in the AHM, known as the partially paired phase, that consists of both paired and unpaired components \cite{2015_Greschner,2017_Zhang}. This phase, as well as other exotic phenomena \cite{Keilmann_2011,2016_Strater}, may be reached by adiabatically ramping Floquet parameters to connect to a target state. Our quantum gas microscope is also well-suited for the microscopic study of entanglement properties of 1D anyons \cite{Islam_2015}. Finally, ultracold atoms may offer a route to engineering non-Abelian anyons such as those in the 1D wire construction for topological quantum computation \cite{Alicea_2011}, as suggested by similarities between the AHM and quasi-1D systems hosting non-Abelian anyons \cite{Liu_2018}, while other approaches include introducing three-body hard-core interactions \cite{Harshman_2020} and controlling anyonic excitations in the Pfaffian state \cite{Sterdyniak_2012,Palm_2021,leonard2022realization}.

\section*{Acknowledgments}
We thank Martin Lebrat, Julian L\'eonard, M. Eric Tai, Nathan Harshman, Fangli Liu, Sebastian Nagies, Hannes Pichler, Luis Santos, and Botao Wang for valuable discussions. This work was supported by the ONR grant No. N000114-18-1-2863. A.E. was supported by the Deutsche Forschungsgemeinschaft (DFG) via the Research Unit FOR 2414 under project No. 277974659. A.V.G. was supported in part by the DoE Quantum Systems Accelerator, NSF QLCI (award No.~OMA-2120757), AFOSR, DoE ASCR Accelerated Research in Quantum Computing program (award No.~DE-SC0020312), DoE ASCR Quantum Testbed Pathfinder program (award No.~DE-SC0019040), NSF PFCQC program, ARO MURI, AFOSR MURI, and DARPA SAVaNT ADVENT.

\bibliography{anyons_qw}

%% file: supplementary.tex

\setcounter{page}{1}

\renewcommand{\thefigure}{S\arabic{figure}}
\renewcommand{\theHfigure}{A\arabic{figure}}
\setcounter{figure}{0}

\section*{Supplementary materials}

\subsection*{Materials and Methods}

\paragraph*{\textbf{Generalized Jordan-Wigner transformation}}
There is an exact correspondence \cite{Keilmann_2011} between the Bose-Hubbard model (BHM) with density-dependent phase ${\mathcal{H}}$ and the anyon-Hubbard model (AHM) 
\begin{equation}
{\mathcal{H}}_A = - J \sum_j \left( {a}_{j}^\dagger {a}_{j-1} + \textrm{h.c.} \right) + \frac{U}{2} \sum_j {n}_j \left( {n}_j - 1 \right),
\end{equation}
describing lattice anyons characterized by creation (annihilation) operators ${a}_j^\dagger$ (${a}_j$) that satisfy the algebra
\begin{align}
{a}_j {a}_k^\dagger - e^{ - i\theta \sigma(j - k)} {a}_k^\dagger {a}_j & = \delta_{jk} \\
{a}_j {a}_k - e^{ i\theta \sigma(j - k)} {a}_k {a}_j & = 0 \\
{a}_j^\dagger {a}_k^\dagger - e^{i\theta \sigma(j - k)} {a}_k^\dagger {a}_j^\dagger & = 0.
\end{align}
In these expressions, $\theta$ is the exchange phase of two particles, and $\sigma$ is the sign function $\sigma(j - k) = +1$, $-1$, $0$ when $j > k$, $j < k$, $j = k$, respectively. Following Ref.\,\cite{Keilmann_2011}, we introduce the generalized Jordan-Wigner transformation
\begin{equation}
{b}_j = {a}_j e^{ - i\theta \sum_{k > j} {n}_k},
\end{equation}
which generates operators ${b}_j$ that satisfy bosonic commutation relations: ${b}_j {b}_k^\dagger - {b}_k^\dagger {b}_j = \delta_{jk}$, ${b}_j {b}_k -{b}_k {b}_j = 0$. Note that the number operators ${n}_j = {a}_j^\dagger {a}_j = {b}_j^\dagger {b}_j$ are left unchanged under this transformation. In terms of the bosonic operators ${b}_j$, the AHM ${\mathcal{H}}_A$ immediately rewrites as Eq.\,\eqref{eq:hamiltonian} of the main text, the BHM with density-dependent phase.

\paragraph*{\textbf{Engineering the effective Hamiltonian}}
We realize the BHM with density-dependent phase by modulating the lattice depth, a technique based on Floquet engineering \cite{Goldman_2014,Bukov_2015,Eckardt_2017}. Following the derivation for fermions in Ref.\,\cite{cardarelli_quantum_2019}, we derive the time-dependent Hamiltonian that effectively realizes Eq.\,\eqref{eq:hamiltonian} of the main text, the BHM with density-dependent phase. In this section, we set the reduced Planck constant $\hbar=1$. Starting from the BHM with site offset $E$, we have
\begin{multline}
{\mathcal{H}}_{\textrm{init}} = -J_0\sum_{j} \left({b}^\dagger_j {b}_{j-1}+\text{h.c.}\right) \\
+ \frac{U_0}{2}\sum_j {n}_j ({n}_j - 1) +E\sum_j j {n}_j,
\end{multline}
where ${b}_j^\dagger$ (${b}_j$) is the bosonic creation (annihilation) operator, $J_0$ is the tunneling amplitude between nearest neighbors, $U_0$ is the on-site, pairwise repulsive interaction energy, and ${n}_j = {b}^\dagger_j {b}_j$ is the particle number operator. Defining ${\mathcal{H}}_{\textrm{int}} = 1/2\sum_j {n}_j ({n}_j - 1)$ and ${\mathcal{H}}_{\textrm{tilt}} = \sum_j j {n}_j$, we perform the unitary transformation to a rotating frame of reference via ${\mathcal{U}}=e^{it \left[ (U_0-U) {\mathcal{H}}_{\textrm{int}} +E {\mathcal{H}}_{\textrm{tilt}} \right]}$, where $U$ is the detuning that corresponds to the effective on-site interaction energy. In this rotating frame, the the system is described by the Hamiltonian $\tilde{\mathcal{H}} = {\mathcal{U}}{\mathcal{H}}_{\textrm{init}}{\mathcal{U}}^\dagger + i \dot{{\mathcal{U}}}{\mathcal{U}}^\dagger$,
\begin{widetext}
\begin{align}
\tilde{\mathcal{H}} &= -J_0 \, e^{it \left[ (U_0 - U) {\mathcal{H}}_{\textrm{int}} +E {\mathcal{H}}_{\textrm{tilt}}  \right]} \left(\sum_j {b}^\dagger_j {b}_{j-1} + \text{h.c.} \right) e^{-it \left[ (U_0-U) {\mathcal{H}}_{\textrm{int}} +E {\mathcal{H}}_{\textrm{tilt}}   \right]} + U {\mathcal{H}}_{\textrm{int}} \\
&= -J_0 \sum_j \prod_{k=j-1,j} e^{it\left[ E k {n}_k+\frac{U_0 - U}{2} {n}_k({n}_k-1)\right]}\left({b}^\dagger_j {b}_{j-1} + \text{h.c.}\right)\prod_{k'=j-1,j}e^{-it\left[ E k' {n}_{k'}+\frac{U_0 - U}{2}{n}_{k'}({n}_{k'}-1)\right]} + U {\mathcal{H}}_{\textrm{int}} \\
&= -J_0 \sum_j \left( {b}^\dagger_j e^{i t \left[ E + (U_0-U) ({n}_j - {n}_{j-1}) \right]} {b}_{j-1} + \text{h.c.} \right) + U {\mathcal{H}}_{\textrm{int}}.
\end{align} 
\end{widetext}

Modulating the lattice depth $V_0$ with amplitude $\delta V \ll V_0$, $V(t) = V_0 + \delta V(t)$ has a proportional effect on the tunneling energy, such that $J(t) = J_0 + \delta J(t)$ \cite{2016_Cardarelli}. In a tilted lattice with site offset $E$ and interaction $U_0$, the energy gaps are $E$, $E+U_0$, and $E-U_0$ for maximum particle number $n_{j,\textrm{max}} = 2$ on site $j$ \cite{Ma_2011}. These gaps can be overcome by modulating the lattice depth at these frequencies with amplitude $\delta V \ll V_0$, such that the tunneling energy becomes  
\begin{align}
J(t) &= J_0 + \sum_{s=1}^3 \delta J_s \cos(\omega_s t + \theta_s) \\
&= J_0 + \sum_{s=1}^3 \frac{\delta J_s}{2} \left( e^{i (\omega_s t + \theta_s)} +  e^{-i (\omega_s t + \theta_s)} \right),
\end{align}
where $\omega_s$ is the frequency and $\theta_s$ is the phase of component $s$. We substitute $J(t)$ into $\tilde{\mathcal{H}}$ to obtain
\begin{widetext}
\begin{multline}
\tilde{\mathcal{H}} = -J_0 \sum_j  \left( {b}^\dagger_j e^{i t \left[ E + (U_0-U) ({n}_j - {n}_{j-1}) \right]} {b}_{j-1} + \textrm{h.c.} \right) \\
- \sum_{s=1}^3 \frac{\delta J_s}{2} \sum_j \left( {b}^\dagger_j e^{i t \left[ E + (U_0-U) ({n}_j - {n}_{j-1}) - \omega_s \right] - i \theta_s} {b}_{j-1} + \text{h.c.} \right) + \frac{U}{2}\sum_j {n}_j ({n}_j - 1).
\end{multline}
\end{widetext}
If we choose the three driving frequencies to be $\omega_1=E$,  $\omega_2=E-(U_0-U)$, $\omega_3=E+(U_0-U)$, the phases associated with the tunneling processes proportional to $\delta J_s$ become time-independent for the targeted density-dependent processes depicted in Fig.\,\ref{fig:protocol}C of the main text. For all other tunneling processes, and in particular those proportional to $J_0$, the tunneling matrix elements average out over time and can therefore be neglected. This rotating-wave approximation is valid as long as $J_0 \ll E, |E \pm (U_0-U)|$. For the targeted processes, the tunneling energy $J$ in the effective Hamiltonian is given by $\delta J_s/2$, which we choose to be equal, while the Peierls phases of the effective tunneling matrix elements directly correspond to driving phases $\theta_s$, which are chosen as depicted in Fig.\,\ref{fig:protocol}C. In this way, the system is described by the time-independent Hamiltonian Eq.\,\eqref{eq:hamiltonian} of the main text.

The effective on-site interaction energy $U$ is a variable determined by the detuning of $\omega_2$ and $\omega_3$ with respect to $E\pm U_0$. For example, $\omega_2 = E-U_0$ and $\omega_3 = E+U_0$ realizes the non-interacting BHM with density-dependent phase; $\omega_2 = E - (U_0-U)$ and $\omega_3 = E + (U_0 -U)$ introduces a repulsive on-site $U$ in the effective Hamiltonian. Note that although the modulation frequencies correspond to the energy gaps for $n_{j,\textrm{max}} = 2$, they can induce further processes involving three or more particles. These processes, however, do not apply to our system of two particles and are generally negligible at low filling \cite{2016_Strater}.

\paragraph*{\textbf{State initialization}}
The experiments begin with a two-dimensional Mott insulator at unity-filling of \textsuperscript{87}Rb in a deep optical lattice ($a=680$ nm) with $V_x = V_y = 45E_\textrm{R}$, where $E_\textrm{R} = h \times 1.24$ kHz is the recoil energy. The initial state is prepared by holographically shaping a laser beam at 760 nm with a digital micromirror device (DMD) in the Fourier plane with respect to the atoms. This DMD allows us to project arbitrary potentials with single lattice site resolution through our diffraction-limited microscope objective \cite{Bakr_2009}, and correct for optical wavefront aberrations in the imaging system \cite{Zupancic_2016}. To prepare the initial state, we use the DMD to optically confine two adjacent columns of atoms along $y$ in the unity-filling shell of the Mott insulator, then reduce $V_x$ before ejecting atoms outside the confinement by projecting a Gaussian repulsive potential. After atoms outside the confinement have been removed, we increase the lattice back to $V_x = 45 E_\textrm{R}$ and turn off the confining potential projected by the DMD. We therefore prepare two columns of atoms along $y$ ready to undergo independent quantum walks along $x$, induced by decreasing $V_x$ while keeping $V_y=45E_\textrm{R}$. We realize $\sim 8 \text{--} 12$ independent quantum walks in decoupled tubes in each experimental run. 

\paragraph*{\textbf{Bose-Hubbard parameters}}
The magnetic field gradient $E$ per lattice site and interaction energy $U_0$ can be simultaneously calibrated using a spectroscopic technique of modulating the lattice depth across a frequency range \cite{Ma_2011}. Starting with a Mott insulator, we apply a magnetic field gradient $E$ per lattice site, lower the lattice depth along $x$ to $V_x = 4 E_\textrm{R}$, the lattice depth used for the quantum walks, and modulate the lattice depth across a frequency range that includes $E-U_0$ and $E+U_0$. Resonances occur at $E-U_0$ and $E+U_0$, when atoms can tunnel to occupied sites, and manifest as decreased probability of singly-occupied sites in fluorescence images due to parity projection. 

We calibrate tunneling energy $J$ by performing single-particle quantum walks in the effective Hamiltonian. Similar to state initialization for two-particle quantum walks, we prepare a single column of atoms along $y$ in a deep optical lattice and apply site offset $E$. Then we lower $V_x$ to $4 E_\textrm{R}$ and modulate the lattice depth at frequency $E'$ by $20\%$ to induce quantum walks along $x$ for various times, averaging over many experimental runs to obtain a density profile to which we fit the distribution from theory \cite{Hartmann_2004},
\begin{equation}
\rho_{|i|}(t) = \left|\mathcal{J}_i\left(\frac{4 J}{\Delta} \sin(\pi \Delta t) \right) \right|^2,
\end{equation}
where $\mathcal{J}_i$ is a Bessel function of the first kind on lattice site $i$ and $\Delta=E'-E$ is the local gradient. This fit also allows us to determine any mismatch between $E$ and $E'$, which leads to Bloch oscillations.

\paragraph*{\textbf{Optimal Floquet parameters}}
Realizing the effective Hamiltonian via Floquet engineering, which in our case is lattice depth modulation, requires fulfillment of a few conditions: frequency of modulation $f_{\textrm{mod}}$ should be low enough not to excite the system to higher bands, but high enough to be well separated from the low energy scale of the effective and initial Hamiltonians \cite{Str_ter_2016,Sun_2020}. For our system, the regime that meets these conditions is $U_0/h < f_{\textrm{mod}} < f_{\textrm{max}}$, where $f_{\textrm{max}} \approx 1$ kHz, determined from bandgap calculations and consistent with results in Ref.\,\cite{Str_ter_2016}. 

Given these conditions, we attribute deviations from theory in Fig.\,\ref{fig:asymmetric_transport} when $U<0$ to Floquet heating and imperfect realization of the effective Hamiltonian. Introducing $U<0$ corresponds to detuning the sidebands such that $ \left[E-(U_0-U)\right]/h \to U_0/h$ and $ \left[E+(U_0-U)\right]/h \to f_{\textrm{max}}$. We see the upper sideband approaches the high frequency limit, leading to excitations to higher bands, while the lower sideband approaches the low frequency limit, reducing viability of the rotating-wave approximation. In addition, heating rates can increase with driving strength \cite{Str_ter_2016}. For us, the optimal driving strength is modulating the lattice depth by $20\%$ of $V_x= 4 E_{\textrm{R}}$, which allows for dynamics to occur within the coherence time of the system while minimizing heating.

\begin{figure*}
\includegraphics{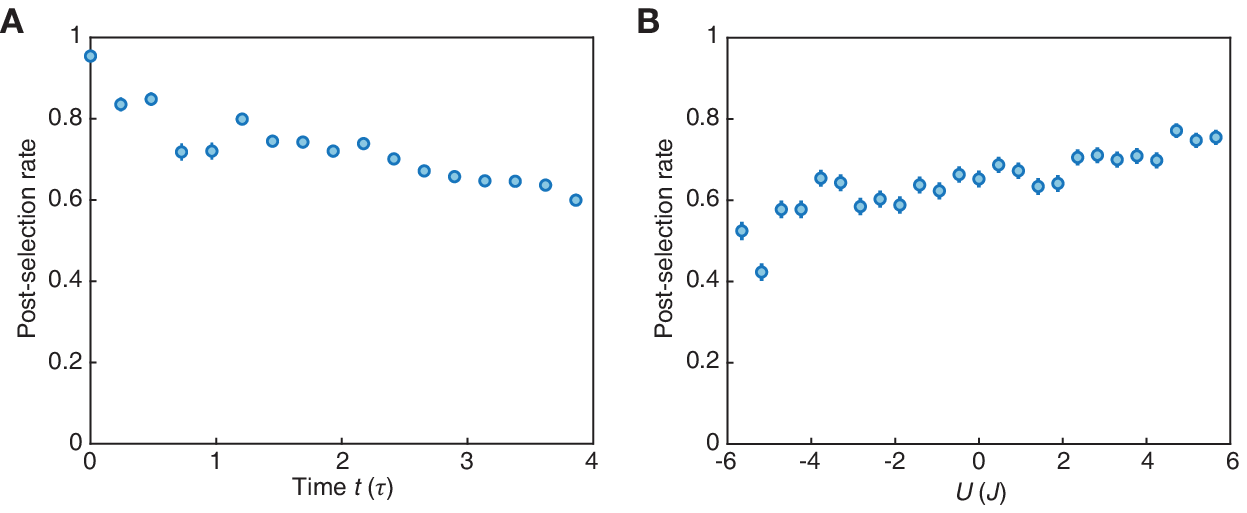}
\caption{\textbf{Post-selection rate.} (A) Post-selection rate for data in Fig.\,\ref{fig:quantum_walk} of the main text, $\theta=\pi$. It decreases with time, signaling slow loss of coherence of the system, but remains reasonably high at $\sim 60\%$ at the end of the experiment time $t\approx 4\tau$. Post-selection rates for other phases $\theta=0,\pi/2$ are comparable. (B) Post-selection rate for data in Fig.\,\ref{fig:asymmetric_transport}E of the main text, $\theta=\pi/2$. It decreases as $U$ approaches the strongly attractive regime because the sidebands correspondingly approach the limits of the optimal Floquet window. The upper sideband $E+(U_0-U)$ approaches resonance with higher bands, leading to Floquet heating, while the lower sideband $E-(U_0-U)$ approaches the low energy scale of the Hamilonian, leading to breakdown of the rotating-wave approximation. Error bars denote the s.e.m.}
\label{fig:sm_postselection}
\end{figure*}

\paragraph*{\textbf{Post-selection}}
When operating within these regimes, the post-selection rate, defined as the proportion of experimental runs with two particles, decreases as time evolves. For a typical two-particle quantum walk with $U=0$, the post-selection rate is $\sim 95\%$ at $t=0$, decreasing to $\sim 60\%$ at $t \approx 4\tau$, where $\tau = 15.0(3)$ ms (Fig.\,\ref{fig:sm_postselection}). As $U$ is detuned from $\sim 6J$ to $\sim -6J$, corresponding to the sidebands approaching the limits of the optimal Floquet regime, $\left[E-(U_0-U)\right]/h \to U_0/h$ and $\left[E+(U_0-U)\right]/h \to f_{\textrm{max}}$, post-selection decreases from $\sim 80\%$ when $U\approx 6J$ to $\sim 50\%$ when $U\approx -6J$ at $t=2.40(5)\tau$. 

\begin{figure*}
\includegraphics{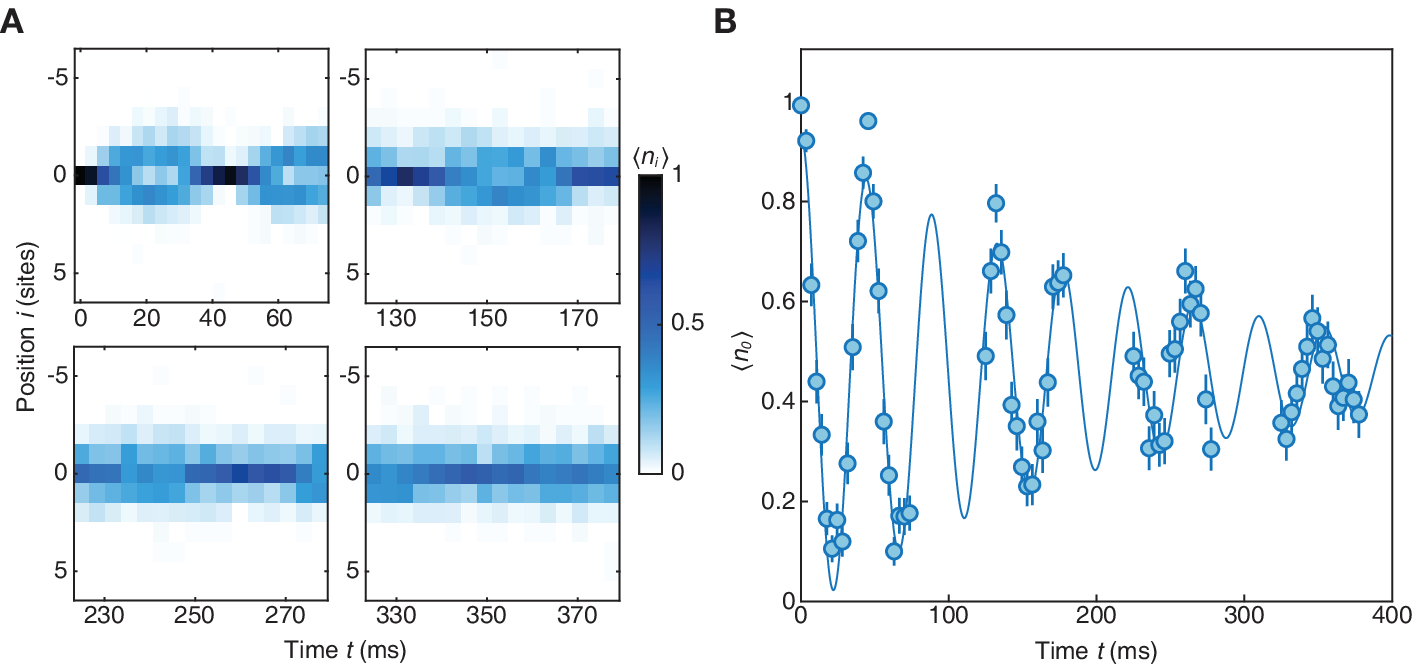}
\caption{\textbf{Coherence time.} (A) Bloch oscillations of a single atom in a tilted lattice. A single atom is initialized on site $i = 0$ in a tilted 1D lattice with energy offset $E/h = 800(2)$ Hz. Then, the lattice depth $V_x = 4 E_{\textrm{R}}$ is modulated by $20\%$ with frequency $\omega_1 = 2\pi \cdot 780$ Hz to restore tunneling, leaving a residual energy offset $E'/h = 20(2)$ Hz that induces Bloch oscillations. Coherent revivals can be seen after $300$ ms evolution time. (B) Density of the initially occupied site $\langle n_0 \rangle$ as a function of time (same data as in A). Solid line is a fit to a damped oscillator function $A \, e^{-2t/\tau_{\textrm{3-tone}}} \cos(2 \omega t) + B$ \cite{Kolovsky_2002}, which yields a coherence time $\tau_{\textrm{3-tone}} = 0.44(5)$ s. Error bars denote the s.e.m.}
\label{fig:sm_coherence}
\end{figure*}

\paragraph*{\textbf{Coherence time}}
We determine the coherence time of the system subject to lattice-depth modulation by performing modulation-induced Bloch oscillations of a single atom (Fig.\,\ref{fig:sm_coherence}). In a tilted lattice with energy offset $E/h = 800(2)$ Hz, we modulate the lattice depth $V_x = 4 E_{\textrm{R}}$ by $20\%$ with frequency $\omega_1 = 2\pi \cdot 780$ Hz to restore tunneling, leaving a residual energy offset $E'/h = 20(2)$ Hz that induces Bloch oscillations. Fitting the density of the initial site, where revivals occur, to a damped oscillator function, we obtain a $1/e$-lifetime of $\tau_{\textrm{3-tone}} = 0.44(5)$ s.

\paragraph*{\textbf{Data analysis}}
Light-assisted collisions prevent us from directly detecting doubly-occupied sites (``parity projection") to obtain diagonal elements $\Gamma_{i,i}$ of the two-particle correlator \cite{Bakr_2009}. We circumvent this limitation using a technique described in Ref.\,\cite{Preiss_2015} that splits pairs of atoms before imaging. At the end of time evolution for the quantum walks, we raise the lattice depth from $V_x=4E_\textrm{R}$ to $V_x=15E_\textrm{R}$ and adiabatically decrease the magnetic field gradient from $E \approx 4 U_0$ to $E \approx 0.5 U_0$ in $250$ ms, passing the $U_0$ resonance at which doubly-occupied sites are converted to atoms on neighboring sites and vice versa ($| ... 2 0 ...\rangle \Leftrightarrow | ... 1 1 ... \rangle$) with $\sim90\%$ fidelity. This procedure amounts to mapping $\Gamma_{i,i+1}$ to $\Gamma_{i,i}$. We perform this detection scheme for half the data set to obtain $\Gamma_{i,i}$ and obtain $\Gamma_{i,i+1}$ directly from the images in the other half. The full correlator $\Gamma_{i,j}$ is obtained by combining the two halves weighted by the number of post-selected realizations.

\paragraph*{\textbf{Fock state populations at short times}}
We follow Ref.\,\cite{Liu_2018} to determine the populations of states $|...1100...\rangle$ and $|...0011...\rangle$ after a short time evolution $t$, when starting from the initial state $|...0110...\rangle$. For convenience, we denote $|+\rangle = |...1100...\rangle$, $|-\rangle = |...0011...\rangle$ and $|0\rangle = |...0110...\rangle$ only in this section. Using a Taylor expansion of the unitary time evolution operator ${\mathcal{U}}_t = e^{-i {\mathcal{H}} t / \hbar}$ about $t=0$, we derive the transition amplitudes toward the Fock states $|\pm\rangle$ up to third order in $t$,
\begin{equation}
\label{eq:amplitude}
a_{\pm}(t) = \langle \pm | {\mathcal{U}}_t | 0 \rangle \simeq \frac{1}{2} \left( \frac{Jt}{\hbar} \right)^2 \left[ 1 + 2 e^{\pm i \theta} \left( 1 - i \, \frac{U t}{3 \hbar} \right) \right].
\end{equation}
In the case $U = 0$, we obtain the analytical prediction $|a_+(t)|^2 = |a_-(t)|^2 = \frac{1}{4} \left( \frac{Jt}{\hbar} \right)^4 \left( 5 + 4 \cos \theta \right)$, which is qualitatively consistent with Fig.\,\ref{fig:interferometer}B of the main text. When $U \neq 0$, $|a_+(t)|^2 \neq |a_-(t)|^2$, and we can rewrite Eq.\,\eqref{eq:amplitude} as
\begin{equation}
\label{eq:amplitude_2}
a_{\pm}(t) \simeq \frac{1}{2} \left( \frac{Jt}{\hbar} \right)^2 \left( 1 + 2 e^{\pm i  \theta - i \frac{U t}{3 \hbar}} \right),
\end{equation}
which is also correct up to third order in $t$. Importantly, Eq.\,\eqref{eq:amplitude_2} allows us to interpret the introduction of on-site interaction $U$ as providing an additional phase $-U t' / \hbar$ with dwell time $t' = t/3$ over the intermediate doubly-occupied site in Fig.\,\ref{fig:asymmetric_transport} of the main text.

\clearpage

\subsection*{Supplementary Text}

\paragraph*{\textbf{Bound states}}

In this section, we derive the solutions for the two-body problem of the BHM with density-dependent phase in an infinite 1D chain, and recall the properties of bound states discussed in Refs.\,\cite{2016_Cardarelli,2017_Zhang,2018_Greschner}. The eigenstates of the Hamiltonian can be expanded over the Fock state basis $| \{n_j\} \rangle$ with site label $j \in \mathbb{Z}$, restricted to $\sum_j n_j = 2$ for the two-particle problem considered here. We use the translational invariance of the BHM to express the eigenstates in terms of the center-of-mass and relative positions. More specifically, we denote the two-particle Fock state $|n,m\rangle$, where $n \in \mathbb{Z}$ is the position of the leftmost particle and $n + m \in \mathbb{Z}$ is the position of the rightmost particle (we restrict $m \in \mathbb{N} = \{0,1,2,...\}$ to be non-negative to account for indistinguishability). Hence, we can expand any eigenstate $|\Psi\rangle$ over this basis,
\begin{equation}
|\Psi\rangle = \sum_{n\in \mathbb{Z}, m \in \mathbb{N}} \tilde c_{nm} \, |n,m\rangle,
\end{equation}
and impose $|\Psi\rangle$ to be an eigenstate of the translation operator by one site, which amounts to expressing $\tilde c_{nm}$ as
\begin{equation}
\label{eq:coefs}
\tilde c_{nm} = e^{ i q \left( n + \frac{m}{2} \right) } c_m,
\end{equation}
where $n + \frac{m}{2}$ is the center-of-mass of the two particles and $q \in [0, 2\pi]$ is the corresponding quasimomentum (in units of the inverse lattice constant). The unknown quantity $c_m$ thus only depends on the relative position $m$. After evaluating ${\mathcal{H}}|\Psi\rangle$, where ${\mathcal{H}}$ is Eq.\,\eqref{eq:hamiltonian}, we look for eigenstates with energy $E_q$ and derive the following linear system for the coefficients $c_m$:
\begin{align}
\label{eq:sys_1}
(\epsilon - u) \, c_0 &= - \sqrt{2} \rho \, c_1, \\
\label{eq:sys_2}
\epsilon \,c_1 &= -\sqrt{2} \rho^* \, c_0 - c_2, \\
\label{eq:sys_3}
\epsilon \, c_m &= - \left( c_{m-1} + c_{m+1} \right),
\hskip5mm m \ge 2,
\end{align}
where we introduced $\gamma = 2 J \cos \left( q/2 \right)$, $\epsilon = E_q/\gamma$, $u = U/\gamma$, $\rho =  2 J \left[  e^{i \left( \frac{q}{2}-\theta \right)} + e^{-i\frac{q}{2} } \right] / \gamma$ and $\rho^*$ its complex conjugate. Note that for generic $\theta$, $\rho$ is a complex number. For $\theta = 0$, $\rho = 1$ is real, while for $\theta = \pi$, $\rho = - i \tan \left( q/2 \right)$ is imaginary. Our goal is to determine the solutions $c_m$ that are physical. Note that Eq.\,\eqref{eq:sys_3} is a recurrence relation whose characteristic polynomial is $x^2 + \epsilon x + 1 = 0$, and is characterized by the quantity $\chi = \epsilon^2 - 4$. A solution $c_m$ is found as a superposition of the two roots $x_\pm$ of this polynomial. Hence, the system \eqref{eq:sys_1}-\eqref{eq:sys_3} has two types of solutions:
\begin{itemize}
\item[--] \textit{Scattering states} made of two counter-propagating plane waves $(c_0, \{c_m =\lambda x_+^m + \mu x_-^m\}_{m \ge 1})$, that correspond to $\chi \le 0$ and that can be obtained for any energy $E_q$ within the continuum:
\begin{equation}
\label{eq:cont}
|E_q| \leq 4 J \cos \left( \frac{q}{2} \right).
\end{equation}
\item[--] \textit{Bound states} of the form $(c_0, \{c_m = x_\pm^m \}_{m \ge 1})$ that decay exponentially with distance and that correspond to $\chi > 0$. These bound states can be grouped into two branches $(\pm)$ lying above (resp. below) the continuum. Their dispersion relation can be obtained for arbitrary $\theta$ and $U$:
\begin{widetext}
\begin{equation}
\label{eq:bound}
E_{q,\pm} / J = - \frac{ U \left[ \cos^2 \frac{q}{2} - \cos^2 \left( \frac{q-\theta}{2} \right) \right] \mp \cos^2 \left( \frac{q-\theta}{2} \right) \sqrt{ U^2 + 16 \left[ 2 \cos^2 \left( \frac{q-\theta}{2} \right) - \cos^2 \frac{q}{2} \right] } }{ 2 \cos^2 \left( \frac{q-\theta}{2} \right) - \cos^2 \frac{q}{2} }.
\end{equation}
\end{widetext}
In the case $U = 0$, this expression simplifies to
\begin{equation}
\label{eq:bound_2}
E_{q,\pm} / J = \pm \frac{ 4 \cos^2 \left( \frac{q-\theta}{2} \right) } { \sqrt{ 2 \cos^2 \left( \frac{q-\theta}{2} \right) - \cos^2 \frac{q}{2} } }.
\end{equation}
Other related expressions for the bound pair dispersion relation can be found in Refs.\,\cite{2017_Zhang,2018_Greschner}. Note that for certain values of $(\theta, U)$, these states do not exist for arbitrary $q$ but only in a restricted interval of quasimomenta. The two branches are symmetric (not symmetric) about $E_q = 0$ for $U = 0$ ($U \neq 0$). The two branches are symmetric (not symmetric) under $q \leftrightarrow -q$ for $\theta = 0, \pi$ ($0 < \theta < \pi$). The latter effect comes from the lack of inversion symmetry of the BHM with density-dependent phase for fractional phase, see Fig.\,\ref{fig:bound_state}A of the main text.
\end{itemize}

\begin{figure}
\includegraphics{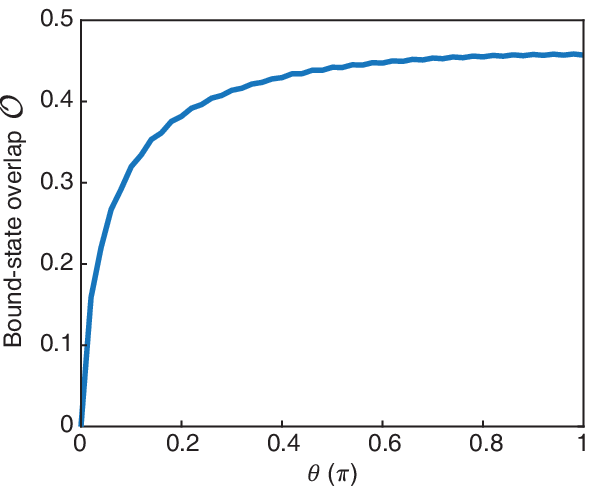}
\caption{\textbf{Analysis of bound states in the AHM.} Overlap $\mathcal{O}$ between the initial state $|...0110...\rangle$ and bound states as a function of $\theta$, $U = 0$.}
\label{fig:sm_overlap}
\end{figure}

\paragraph*{\textbf{Bound states in the expansion dynamics}}
In our experiments, we observe the expansion dynamics of the initial state $|\Psi_0\rangle = |...0110...\rangle$. Such a spatially-localized state projects onto all possible quasimomentum components. Remarkably, the dispersion relation of bound states found for $U = 0$ becomes flatter as $\theta$ grows from $0$ to $\pi$ so that the group velocity decreases, explaining the slowing down of the expansion of nearby particles shown in Fig.\,\ref{fig:bound_state} of the main text. We perform numerical simulations to complement our analysis of the bound-state dynamics in Fig.\,\ref{fig:bound_state} of the main text, using a 1D chain of 80 sites with periodic boundary conditions, without on-site interaction $U$ and without residual tilt in the BHM with density-dependent phase. First, we compute the overlap of the initial state $|\Psi_0\rangle$ with the family of bound states determined above via exact diagonalization. We define the overlap between the initial state $|\Psi_0\rangle$ and the bound states as
\begin{equation}
\mathcal{O} = \sum_{ \ell \in \Lambda} \left| \langle \Phi_\ell | \Psi_0 \rangle \right|^2,
\end{equation}
where $\Lambda$ is the set of indices labelling the bound eigenstates $|\Phi_\ell\rangle$. As shown in Fig.\,\ref{fig:sm_overlap}, overlap with bound states is zero when $\theta = 0$ (as there are no bound states in this case) and rises to $> 40\%$ as $\theta$ grows to $\pi$.

\begin{figure*}
\includegraphics{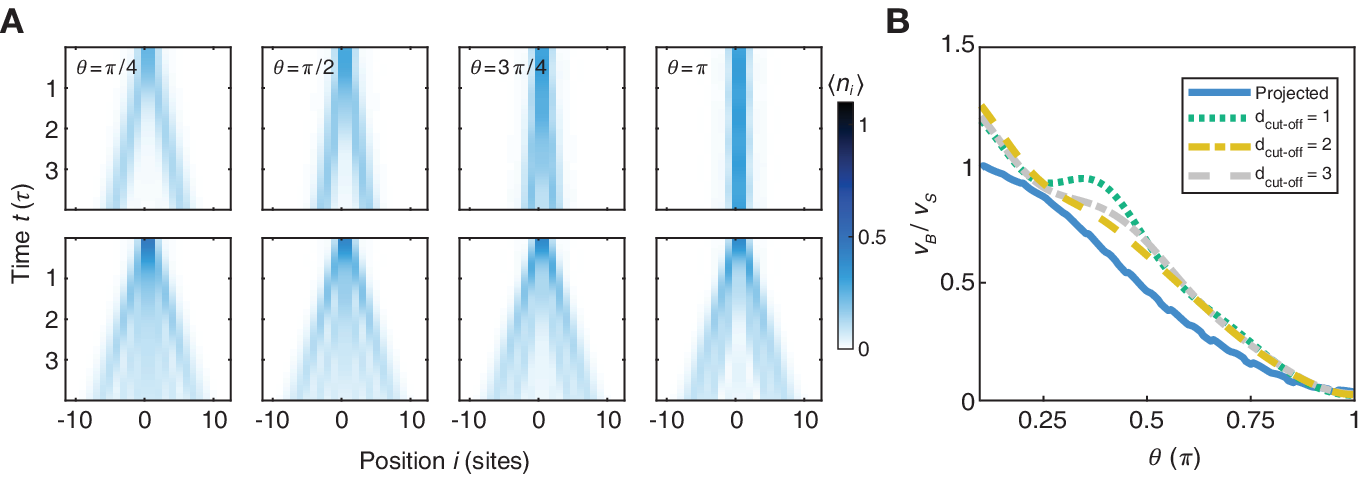}
\caption{\textbf{Contribution of bound and scattering components to expansion dynamics.} (A) Numerical evolution of the density profiles after projecting onto the family of bound states (top row) and scattering states (bottom row), for various statistical phases $\theta$. (B) We extract the spreading velocity $v_B$ ($v_S$) for the bound (scattering) component by fitting the evolution of the root-mean-square size at time $t > \tau$. We compare the ratio $v_B/ v_S$ (solid line) to the analysis in Fig.\,\ref{fig:bound_state} of the main text, where we separate the wavefunction components based on the relative distance $d$ between the two particles, for various cut-off distances $d_{\textrm{cut-off}}$ (dashed lines). }
\label{fig:sm_bound_vs_scattering}
\end{figure*}

Then, we show that conditioning the density profiles on the relative distance $d_{\textrm{cut-off}}$ between the two particles approximately distinguishes the scattering and bound components for our initial state $|\Psi_0\rangle$. While these two components cannot be prepared or detected separately in our experiment, we show in Fig.\,\ref{fig:sm_bound_vs_scattering}A the expected dynamics of the bound (top row) and scattering (bottom row) components for various statistical phases $\theta$. These are obtained by projecting the evolved quantum state $|\psi(t)\rangle = e^{-i\mathcal{H}t} |\Psi_0\rangle$, where $\mathcal{H}$ is Eq.\,\eqref{eq:hamiltonian}, onto the two corresponding Hilbert subspaces. 

By fitting the evolution of the root-mean-square size for both components, we extract the spreading velocities $v_B$ and $v_S$ for the bound and scattering components, respectively. We show the ratio of these two velocities $v_B/v_S$ in Fig.\,\ref{fig:sm_bound_vs_scattering}B (solid line), for statistical phases $\theta > 0.1\,\pi$. When $\theta \simeq 0$, we find that the bound component vanishes almost completely, since our initial state mostly projects onto scattering states at this value of $\theta$ (Fig.\,\ref{fig:sm_overlap}). This component also gradually becomes more spatially extended, which makes the analysis more sensitive to finite-size effects. Fluctuations of the ratio of spreading velocities $v_B/v_S$ for $\theta > 0.1\,\pi$ also indicate finite-size effects. In a finite-size system, the available quasimomenta $q$ are discretized so that the number of bound states entering the decomposition of $|\Psi_0\rangle$ changes by discrete steps as $\theta$ varies over $[0, \pi]$. We show in Fig.\,\ref{fig:sm_bound_vs_scattering} the results of the analysis made in Fig.\,\ref{fig:bound_state} of the main text, for various cut-off distances $d_{\textrm{cut-off}} = 1$, $2$, $3$ sites (dashed lines). We find that the different curves show a similar trend, which confirms the validity of the method described in the main text and based on $d_{\textrm{cut-off}} = 2$.

\begin{figure*}
\includegraphics{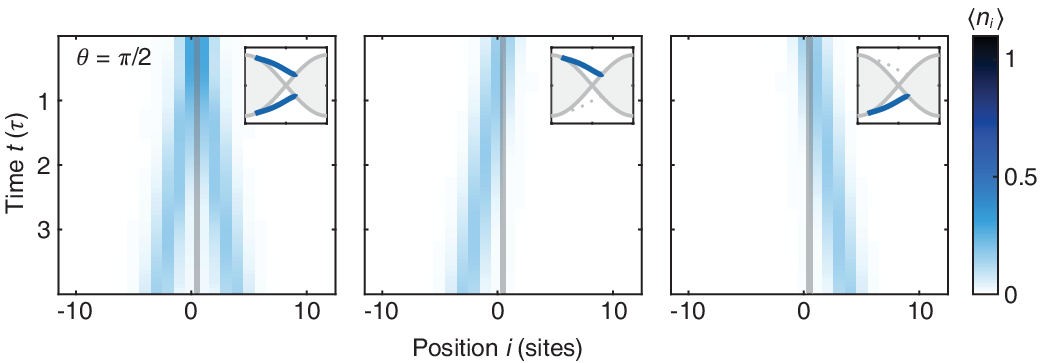}
\caption{\textbf{Contribution of the upper and lower bound-state branches.} The density profile of the bound-state component for $\theta = \pi/2$ (left) can be further decomposed numerically by projecting the quantum state $|\psi(t)\rangle$ onto the bound states of the upper branch (middle) and the lower branch (right). The grey solid line at the center is a guide to the eye. The preferred direction of propagation for each branch is consistent with the sign of the group velocity $\textrm{d}E_q/\textrm{d}q$. The overall density evolution is left-right symmetric because the two branches participate equally in the decomposition of the initial state $|...0110...\rangle$.}
\label{fig:sm_left_right}
\end{figure*}

Next, we discuss the influence of each bound-state branch on the evolution of the density profile. As shown in Fig.\,\ref{fig:bound_state}A, for $0 < \theta < \pi$, the lower and upper branches individually show a preferred direction of propagation, given by the sign of the group velocity $\textrm{d}E_q/\textrm{d}q$. Our initial state $|\Psi_0\rangle$ projects onto scattering states and both branches of bound states. Numerically, we find that for each value of $q$, the bound states of the lower and the upper branches contribute with equal weight and show exactly opposite group velocities. Therefore, the internal cones in the density profiles of Fig.\,\ref{fig:bound_state}B of the main text appear symmetric about the center. In Fig.\,\ref{fig:sm_left_right}B and S5C, we show that each bound-state branch indeed shows chiral propagation, as it corresponds to one edge of the internal cone. Interestingly, the fact that the density profile from our initial Fock state shows symmetric propagation can be related to a dynamical symmetry discussed in Ref.\,\cite{Liu_2018}.

\begin{figure*}
\includegraphics{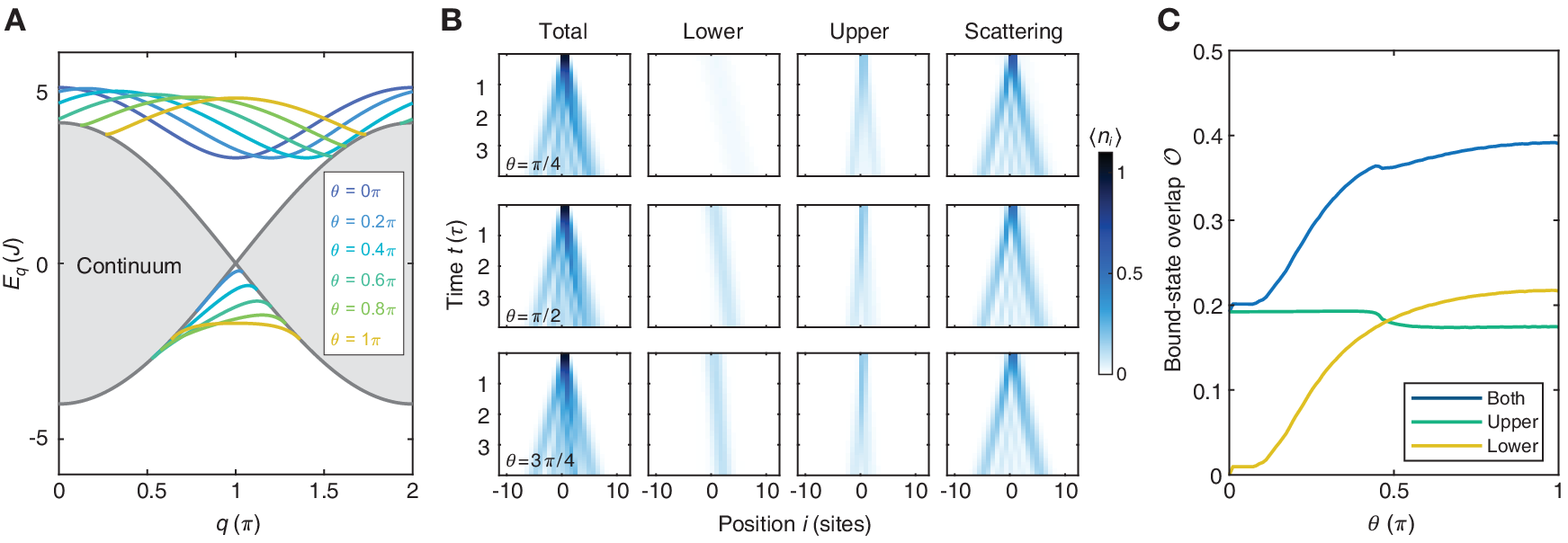}
\caption{\textbf{Bound states with on-site interaction $U$.} All numerical results shown in this figure have been obtained for the case $U = 3J > 0$. (A) Two-particle spectrum $E_q$ as a function of the center-of-mass quasimomentum $q$, for various statistical phases $\theta$. The continuum is present for any $\theta$. Unlike the $U=0$ case, the upper branch is no longer the reflection of the lower branch about $E_q = 0$, and the lower branch shows a more pronounced directionality (towards the right) than the upper branch. Plots for negative values of $\theta$ are not shown as they are simply obtained by reflecting the curves with positive $\theta$ across $q = \pi$. (B) Density profiles from two-particle quantum walks (first column). The three rows correspond to different values of $\theta$, while the columns respectively show the contributions from the lower branch (second column), the upper branch (third column), and the scattering states (fourth column). By fitting the center-of-mass trajectory from the first to the fourth column, we extract center-of-mass velocities respectively equal to $(0.23, 1.25, 0.02, 0.16)$ [$\theta = \pi/4$], $(0.23, 0.84, -0.07, 0.11)$ [$\theta = \pi/2$], and $(0.12, 0.46, -0.10, 0.06)$ [$\theta = 3\pi/4$] sites$/\tau$. (C) Overlap between the initial state $|...0110...\rangle$ and the upper branch (red solid line) and lower branch (yellow solid line). The blue solid line is the sum of the two other lines and gives the total overlap with bound states.}
\label{fig:sm_interactions}
\end{figure*}

Finally, we discuss how introducing on-site interaction $U$ affects the two-particle spectrum of the BHM with density-dependent phase. In particular, the modified spectrum is sufficient to explain qualitatively the emergence of asymmetric transport in Fig.\,\ref{fig:asymmetric_transport} of the main text, and thus complements the interferometric picture developed in the main text. Numerically, we focus on the value $U = 3\,J > 0$, which describes well the data of Fig.\,\ref{fig:asymmetric_transport}, yet our results remain qualitatively unchanged for other values of $U$. In Fig.\,\ref{fig:sm_interactions}A, we show the two-particle spectrum $E_q$ derived for an infinite chain. Compared to the spectrum plotted in Fig.\,\ref{fig:bound_state}A for the case $U = 0$, the lower and upper bound-state branches are no longer symmetric abount $E_q=0$: the lower branch has a positive group velocity $\textrm{d}E_q / \textrm{d} q$, which is on average larger than for the upper branch (in absolute value), and thus suggests rightward transport. Note that, contrary to the case $U = 0$ shown in Fig.\,\ref{fig:bound_state}A, there exist bound states -- albeit only in the upper branch (dark blue curve in Fig.\,\ref{fig:sm_interactions}A) -- even for $\theta = 0$. In the case of large $|U|$, this branch is clearly separated from the continuum and describes the dispersion relation of attractively- (repulsively-) bound pairs when $U < 0$ ($U > 0$). 

We further confirm the asymmetry in the group velocities by decomposing the two-particle quantum walks into its various components for different values of $\theta$, shown in Fig.\,\ref{fig:sm_interactions}B. Next to the total density profile (first column), we show the density profiles obtained after projecting the wavefunction respectively onto the lower branch, the upper branch, and the scattering states that contain the rest of the spectrum. Note that the total density profile is not obtained by merely summing these three components, as these components can interfere with one another. We deduce from Fig.\,\ref{fig:sm_interactions}B that the lower branch shows more asymmetric transport than the upper branch. This asymmetry is maximal for small $\theta$, as can be anticipated from the two-particle spectrum shown in Fig.\,\ref{fig:sm_interactions}A. We quantify this asymmetry by measuring the center-of-mass velocity for each component (see caption of Fig.\,\ref{fig:sm_interactions}B). While the lower branch explains a significant part of the left-right asymmetry of the walk, we also note significant asymmetry from the scattering-state component. Furthermore, we observe from Fig.\,\ref{fig:sm_interactions}B that the weight of the lower branch in the dynamics strongly depends on $\theta$ and is smallest at small $\theta$. Quantitatively, we extract the overlap between our initial state $|\Psi_0\rangle$ and the lower and upper branches separately as a function of $\theta$, shown in Fig.\,\ref{fig:sm_interactions}C. The $\theta$-dependence of the group velocity together with the overlap with the lower branch explain why we observe maximally asymmetric transport for a statistical phase $\theta = \pi/2$ out the three values of $\theta$ shown here, in agreement with Fig.\,\ref{fig:asymmetric_transport}D of the main text.

\begin{figure*}
\includegraphics{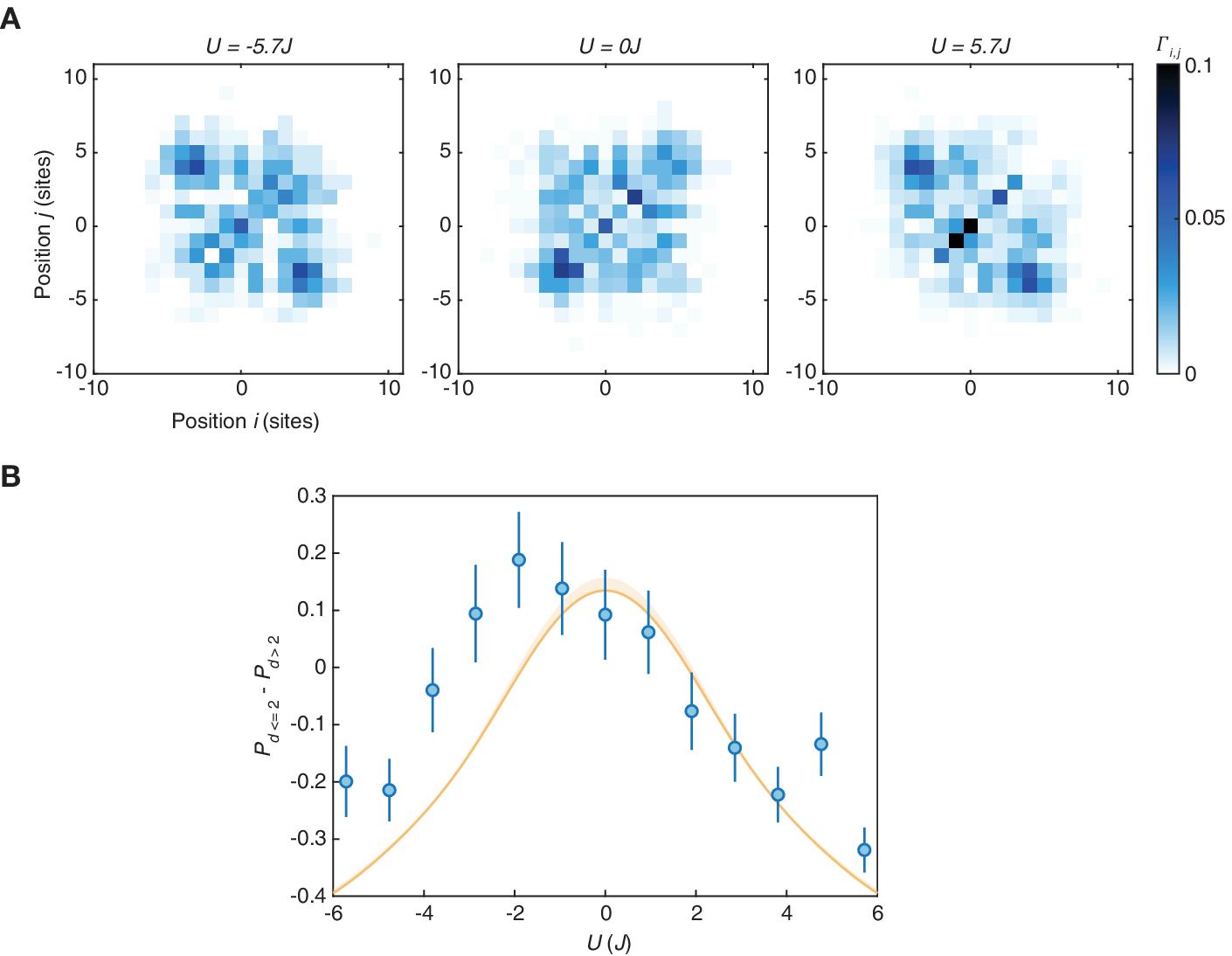}
\caption{\textbf{Bosons with tunable interaction $U$.} (A) Density correlators $\Gamma_{i,j}$ for bosons ($\theta = 0$) at $t=2.40(5)\tau$, from strongly attractive ($U=-5.7(4)J$) to strongly repulsive ($U=5.7(4)J$) effective on-site interaction energy $U$, showing the fermionization of bosons for strong interactions. (B) Bunching parameter $P_{d \le 2} - P_{d >2}$ as a function of the effective on-site energy $U$. The yellow line indicates the results from exact diagonalization, with the shaded region accounting for uncertainty in the tilt calibration. The deviation from theory on the attractive side is likely due to enhanced Floquet heating for our highest shaking frequencies. Error bars denote the s.e.m.}
\label{fig:sm_bosons}
\end{figure*}

\paragraph*{\textbf{Bosons with tunable interactions $U$}}
As a complement to the experimental data reported in the main text, we demonstrate in Fig.\,\ref{fig:sm_bosons} our ability to tune the effective on-site interaction $U$ without relying on Feshbach resonance. In Fig.\,\ref{fig:sm_bosons}A, we focus on the case of bosons ($\theta = 0$) and observe fermionization for strongly attractive and strongly repulsive interactions. Fermionization is manifested in the density correlator $\Gamma_{i,j}$ by weights in the anti-diagonal and indicative of anti-bunching \cite{Preiss_2015}, as opposed to the bunching behavior visible for $U = 0$. We further confirm this effect in Fig.\,\ref{fig:sm_bosons}B by plotting the difference $P_{d \le 2} - P_{d >2}$, where $P_{d \le 2}$ ($ P_{d >2}$) is the probability to find the particles at relative distance $d \le 2$ ($d >2$). The residual asymmetry in Fig.\,\ref{fig:sm_bosons}B is likely due to Floquet heating for our largest shaking frequencies.

\begin{figure*}
\includegraphics{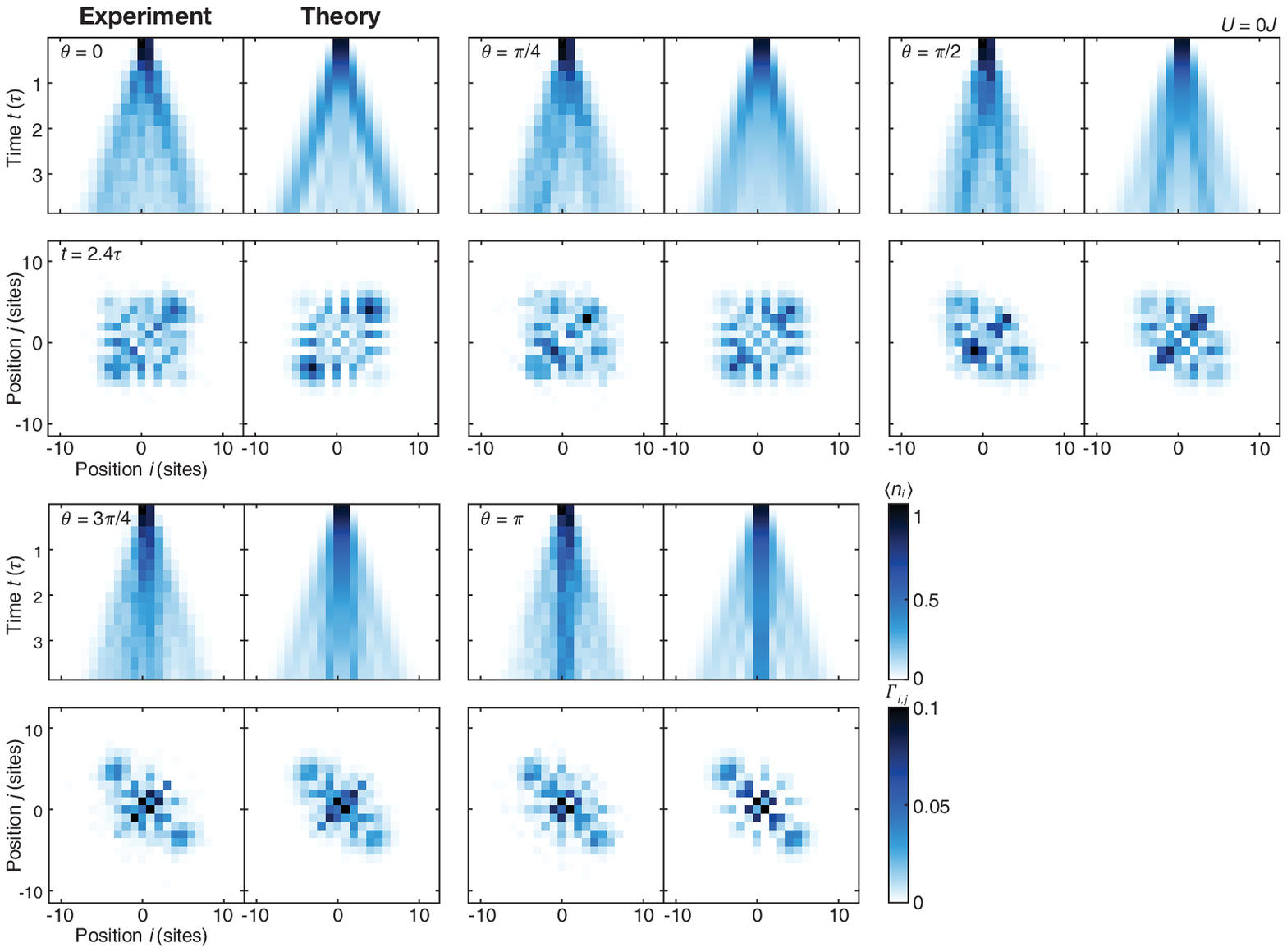}
\caption{\textbf{Full data set for quantum walks of two anyons, $U=0$.} Data for $\theta=0$, $\pi/2$, and $\pi$ are the same as in Fig.\,\ref{fig:quantum_walk} of the main text.}
\label{fig:sm_complete_qw}
\end{figure*}

\paragraph*{\textbf{Quantum walks of two anyons, $U=0$}}
We show the complete data set for quantum walks of two anyons in Fig.\,\ref{fig:sm_complete_qw}, where the density profiles and two-particle correlations are also shown for $\theta=\pi/4,\, 3\pi/4$ in addition to data in Fig.\,\ref{fig:quantum_walk} of the main text.

%% file: anyons_qw.bbl
\begin{thebibliography}{65}%
\makeatletter
\providecommand \@ifxundefined [1]{%
 \@ifx{#1\undefined}
}%
\providecommand \@ifnum [1]{%
 \ifnum #1\expandafter \@firstoftwo
 \else \expandafter \@secondoftwo
 \fi
}%
\providecommand \@ifx [1]{%
 \ifx #1\expandafter \@firstoftwo
 \else \expandafter \@secondoftwo
 \fi
}%
\providecommand \natexlab [1]{#1}%
\providecommand \enquote  [1]{``#1''}%
\providecommand \bibnamefont  [1]{#1}%
\providecommand \bibfnamefont [1]{#1}%
\providecommand \citenamefont [1]{#1}%
\providecommand \href@noop [0]{\@secondoftwo}%
\providecommand \href [0]{\begingroup \@sanitize@url \@href}%
\providecommand \@href[1]{\@@startlink{#1}\@@href}%
\providecommand \@@href[1]{\endgroup#1\@@endlink}%
\providecommand \@sanitize@url [0]{\catcode `\\12\catcode `\$12\catcode
  `\&12\catcode `\#12\catcode `\^12\catcode `\_12\catcode `\%12\relax}%
\providecommand \@@startlink[1]{}%
\providecommand \@@endlink[0]{}%
\providecommand \url  [0]{\begingroup\@sanitize@url \@url }%
\providecommand \@url [1]{\endgroup\@href {#1}{\urlprefix }}%
\providecommand \urlprefix  [0]{URL }%
\providecommand \Eprint [0]{\href }%
\providecommand \doibase [0]{https://doi.org/}%
\providecommand \selectlanguage [0]{\@gobble}%
\providecommand \bibinfo  [0]{\@secondoftwo}%
\providecommand \bibfield  [0]{\@secondoftwo}%
\providecommand \translation [1]{[#1]}%
\providecommand \BibitemOpen [0]{}%
\providecommand \bibitemStop [0]{}%
\providecommand \bibitemNoStop [0]{.\EOS\space}%
\providecommand \EOS [0]{\spacefactor3000\relax}%
\providecommand \BibitemShut  [1]{\csname bibitem#1\endcsname}%
\let\auto@bib@innerbib\@empty
\bibitem [{\citenamefont {Leinaas}\ and\ \citenamefont
  {Myrheim}(1977)}]{leinaas_theory_1977}%
  \BibitemOpen
  \bibfield  {author} {\bibinfo {author} {\bibfnamefont {J.~M.}\ \bibnamefont
  {Leinaas}}\ and\ \bibinfo {author} {\bibfnamefont {J.}~\bibnamefont
  {Myrheim}},\ }\href {https://doi.org/10.1007/BF02727953} {\bibfield
  {journal} {\bibinfo  {journal} {Il Nuovo Cimento B}\ }\textbf {\bibinfo
  {volume} {37}},\ \bibinfo {pages} {1} (\bibinfo {year} {1977})}\BibitemShut
  {NoStop}%
\bibitem [{\citenamefont {Wilczek}(1982)}]{Wilczek_1982}%
  \BibitemOpen
  \bibfield  {author} {\bibinfo {author} {\bibfnamefont {F.}~\bibnamefont
  {Wilczek}},\ }\href {https://doi.org/10.1103/PhysRevLett.49.957} {\bibfield
  {journal} {\bibinfo  {journal} {Phys. Rev. Lett.}\ }\textbf {\bibinfo
  {volume} {49}},\ \bibinfo {pages} {957} (\bibinfo {year} {1982})}\BibitemShut
  {NoStop}%
\bibitem [{\citenamefont {Greiter}\ and\ \citenamefont
  {Wilczek}(2022)}]{Greiter_2022}%
  \BibitemOpen
  \bibfield  {author} {\bibinfo {author} {\bibfnamefont {M.}~\bibnamefont
  {Greiter}}\ and\ \bibinfo {author} {\bibfnamefont {F.}~\bibnamefont
  {Wilczek}},\ }\href@noop {} {\bibinfo {title} {Fractional statistics}}
  (\bibinfo {year} {2022}),\ \Eprint {https://arxiv.org/abs/2210.02530}
  {arXiv:2210.02530 [cond-mat.str-el]} \BibitemShut {NoStop}%
\bibitem [{\citenamefont {Halperin}(1984)}]{Halperin_1984}%
  \BibitemOpen
  \bibfield  {author} {\bibinfo {author} {\bibfnamefont {B.~I.}\ \bibnamefont
  {Halperin}},\ }\href {https://doi.org/10.1103/PhysRevLett.52.1583} {\bibfield
   {journal} {\bibinfo  {journal} {Phys. Rev. Lett.}\ }\textbf {\bibinfo
  {volume} {52}},\ \bibinfo {pages} {1583} (\bibinfo {year}
  {1984})}\BibitemShut {NoStop}%
\bibitem [{\citenamefont {Arovas}\ \emph {et~al.}(1984)\citenamefont {Arovas},
  \citenamefont {Schrieffer},\ and\ \citenamefont {Wilczek}}]{Arovas_1984}%
  \BibitemOpen
  \bibfield  {author} {\bibinfo {author} {\bibfnamefont {D.}~\bibnamefont
  {Arovas}}, \bibinfo {author} {\bibfnamefont {J.~R.}\ \bibnamefont
  {Schrieffer}},\ and\ \bibinfo {author} {\bibfnamefont {F.}~\bibnamefont
  {Wilczek}},\ }\href {https://doi.org/10.1103/PhysRevLett.53.722} {\bibfield
  {journal} {\bibinfo  {journal} {Phys. Rev. Lett.}\ }\textbf {\bibinfo
  {volume} {53}},\ \bibinfo {pages} {722} (\bibinfo {year} {1984})}\BibitemShut
  {NoStop}%
\bibitem [{\citenamefont {Bartolomei}\ \emph {et~al.}(2020)\citenamefont
  {Bartolomei}, \citenamefont {Kumar}, \citenamefont {Bisognin}, \citenamefont
  {Marguerite}, \citenamefont {Berroir}, \citenamefont {Bocquillon},
  \citenamefont {Pla{\c{c} }ais}, \citenamefont {Cavanna}, \citenamefont
  {Dong}, \citenamefont {Gennser}, \citenamefont {Jin},\ and\ \citenamefont
  {F{\`{e}}ve}}]{Bartolomei_2020}%
  \BibitemOpen
  \bibfield  {author} {\bibinfo {author} {\bibfnamefont {H.}~\bibnamefont
  {Bartolomei}}, \bibinfo {author} {\bibfnamefont {M.}~\bibnamefont {Kumar}},
  \bibinfo {author} {\bibfnamefont {R.}~\bibnamefont {Bisognin}}, \bibinfo
  {author} {\bibfnamefont {A.}~\bibnamefont {Marguerite}}, \bibinfo {author}
  {\bibfnamefont {J.-M.}\ \bibnamefont {Berroir}}, \bibinfo {author}
  {\bibfnamefont {E.}~\bibnamefont {Bocquillon}}, \bibinfo {author}
  {\bibfnamefont {B.}~\bibnamefont {Pla{\c{c} }ais}}, \bibinfo {author}
  {\bibfnamefont {A.}~\bibnamefont {Cavanna}}, \bibinfo {author} {\bibfnamefont
  {Q.}~\bibnamefont {Dong}}, \bibinfo {author} {\bibfnamefont {U.}~\bibnamefont
  {Gennser}}, \bibinfo {author} {\bibfnamefont {Y.}~\bibnamefont {Jin}},\ and\
  \bibinfo {author} {\bibfnamefont {G.}~\bibnamefont {F{\`{e}}ve}},\ }\href
  {https://doi.org/10.1126/science.aaz5601} {\bibfield  {journal} {\bibinfo
  {journal} {Science}\ }\textbf {\bibinfo {volume} {368}},\ \bibinfo {pages}
  {173} (\bibinfo {year} {2020})}\BibitemShut {NoStop}%
\bibitem [{\citenamefont {Nakamura}\ \emph {et~al.}(2020)\citenamefont
  {Nakamura}, \citenamefont {Liang}, \citenamefont {Gardner},\ and\
  \citenamefont {Manfra}}]{Nakamura_2020}%
  \BibitemOpen
  \bibfield  {author} {\bibinfo {author} {\bibfnamefont {J.}~\bibnamefont
  {Nakamura}}, \bibinfo {author} {\bibfnamefont {S.}~\bibnamefont {Liang}},
  \bibinfo {author} {\bibfnamefont {G.~C.}\ \bibnamefont {Gardner}},\ and\
  \bibinfo {author} {\bibfnamefont {M.~J.}\ \bibnamefont {Manfra}},\ }\href
  {https://doi.org/10.1038/s41567-020-1019-1} {\bibfield  {journal} {\bibinfo
  {journal} {Nat. Phys.}\ }\textbf {\bibinfo {volume} {16}},\ \bibinfo {pages}
  {931} (\bibinfo {year} {2020})}\BibitemShut {NoStop}%
\bibitem [{\citenamefont {Kitaev}(2003)}]{Kitaev_2003}%
  \BibitemOpen
  \bibfield  {author} {\bibinfo {author} {\bibfnamefont {A.}~\bibnamefont
  {Kitaev}},\ }\href {https://doi.org/10.1016/s0003-4916(02)00018-0} {\bibfield
   {journal} {\bibinfo  {journal} {Ann. Phys. (NY)}\ }\textbf {\bibinfo
  {volume} {303}},\ \bibinfo {pages} {2} (\bibinfo {year} {2003})}\BibitemShut
  {NoStop}%
\bibitem [{\citenamefont {Bravyi}(2006)}]{Bravyi_2006}%
  \BibitemOpen
  \bibfield  {author} {\bibinfo {author} {\bibfnamefont {S.}~\bibnamefont
  {Bravyi}},\ }\href {https://doi.org/10.1103/PhysRevA.73.042313} {\bibfield
  {journal} {\bibinfo  {journal} {Phys. Rev. A}\ }\textbf {\bibinfo {volume}
  {73}},\ \bibinfo {pages} {042313} (\bibinfo {year} {2006})}\BibitemShut
  {NoStop}%
\bibitem [{\citenamefont {Nayak}\ \emph {et~al.}(2008)\citenamefont {Nayak},
  \citenamefont {Simon}, \citenamefont {Stern}, \citenamefont {Freedman},\ and\
  \citenamefont {Sarma}}]{Nayak_2008}%
  \BibitemOpen
  \bibfield  {author} {\bibinfo {author} {\bibfnamefont {C.}~\bibnamefont
  {Nayak}}, \bibinfo {author} {\bibfnamefont {S.~H.}\ \bibnamefont {Simon}},
  \bibinfo {author} {\bibfnamefont {A.}~\bibnamefont {Stern}}, \bibinfo
  {author} {\bibfnamefont {M.}~\bibnamefont {Freedman}},\ and\ \bibinfo
  {author} {\bibfnamefont {S.~D.}\ \bibnamefont {Sarma}},\ }\href
  {https://doi.org/10.1103/revmodphys.80.1083} {\bibfield  {journal} {\bibinfo
  {journal} {Rev. Mod. Phys.}\ }\textbf {\bibinfo {volume} {80}},\ \bibinfo
  {pages} {1083} (\bibinfo {year} {2008})}\BibitemShut {NoStop}%
\bibitem [{\citenamefont {Clarke}\ \emph {et~al.}(2013)\citenamefont {Clarke},
  \citenamefont {Alicea},\ and\ \citenamefont {Shtengel}}]{Clarke_2013}%
  \BibitemOpen
  \bibfield  {author} {\bibinfo {author} {\bibfnamefont {D.~J.}\ \bibnamefont
  {Clarke}}, \bibinfo {author} {\bibfnamefont {J.}~\bibnamefont {Alicea}},\
  and\ \bibinfo {author} {\bibfnamefont {K.}~\bibnamefont {Shtengel}},\ }\href
  {https://doi.org/10.1038/ncomms2340} {\bibfield  {journal} {\bibinfo
  {journal} {Nat. Commun.}\ }\textbf {\bibinfo {volume} {4}},\ \bibinfo {pages}
  {1348} (\bibinfo {year} {2013})}\BibitemShut {NoStop}%
\bibitem [{\citenamefont {Andersen}\ \emph {et~al.}(2022)\citenamefont
  {Andersen}, \citenamefont {Lensky}, \citenamefont {Kechedzhi}, \citenamefont
  {Drozdov}, \citenamefont {Bengtsson}, \citenamefont {Hong}, \citenamefont
  {Morvan}, \citenamefont {Mi}, \citenamefont {Opremcak}, \citenamefont
  {Acharya}, \citenamefont {Allen}, \citenamefont {Ansmann}, \citenamefont
  {Arute}, \citenamefont {Arya}, \citenamefont {Asfaw}, \citenamefont
  {Atalaya}, \citenamefont {Babbush}, \citenamefont {Bacon}, \citenamefont
  {Bardin}, \citenamefont {Bortoli}, \citenamefont {Bourassa}, \citenamefont
  {Bovaird}, \citenamefont {Brill}, \citenamefont {Broughton}, \citenamefont
  {Buckley}, \citenamefont {Buell}, \citenamefont {Burger}, \citenamefont
  {Burkett}, \citenamefont {Bushnell}, \citenamefont {Chen}, \citenamefont
  {Chiaro}, \citenamefont {Chik}, \citenamefont {Chou}, \citenamefont {Cogan},
  \citenamefont {Collins}, \citenamefont {Conner}, \citenamefont {Courtney},
  \citenamefont {Crook}, \citenamefont {Curtin}, \citenamefont {Debroy},
  \citenamefont {Barba}, \citenamefont {Demura}, \citenamefont {Dunsworth},
  \citenamefont {Eppens}, \citenamefont {Erickson}, \citenamefont {Faoro},
  \citenamefont {Farhi}, \citenamefont {Fatemi}, \citenamefont {Ferreira},
  \citenamefont {Burgos}, \citenamefont {Forati}, \citenamefont {Fowler},
  \citenamefont {Foxen}, \citenamefont {Giang}, \citenamefont {Gidney},
  \citenamefont {Gilboa}, \citenamefont {Giustina}, \citenamefont {Gosula},
  \citenamefont {Dau}, \citenamefont {Gross}, \citenamefont {Habegger},
  \citenamefont {Hamilton}, \citenamefont {Hansen}, \citenamefont {Harrigan},
  \citenamefont {Harrington}, \citenamefont {Heu}, \citenamefont {Hilton},
  \citenamefont {Hoffmann}, \citenamefont {Huang}, \citenamefont {Huff},
  \citenamefont {Huggins}, \citenamefont {Ioffe}, \citenamefont {Isakov},
  \citenamefont {Iveland}, \citenamefont {Jeffrey}, \citenamefont {Jiang},
  \citenamefont {Jones}, \citenamefont {Juhas}, \citenamefont {Kafri},
  \citenamefont {Khattar}, \citenamefont {Khezri}, \citenamefont {Kieferová},
  \citenamefont {Kim}, \citenamefont {Kitaev}, \citenamefont {Klimov},
  \citenamefont {Klots}, \citenamefont {Korotkov}, \citenamefont {Kostritsa},
  \citenamefont {Kreikebaum}, \citenamefont {Landhuis}, \citenamefont {Laptev},
  \citenamefont {Lau}, \citenamefont {Laws}, \citenamefont {Lee}, \citenamefont
  {Lee}, \citenamefont {Lester}, \citenamefont {Lill}, \citenamefont {Liu},
  \citenamefont {Locharla}, \citenamefont {Lucero}, \citenamefont {Malone},
  \citenamefont {Martin}, \citenamefont {McClean}, \citenamefont {McCourt},
  \citenamefont {McEwen}, \citenamefont {Miao}, \citenamefont {Mieszala},
  \citenamefont {Mohseni}, \citenamefont {Montazeri}, \citenamefont {Mount},
  \citenamefont {Movassagh}, \citenamefont {Mruczkiewicz}, \citenamefont
  {Naaman}, \citenamefont {Neeley}, \citenamefont {Neill}, \citenamefont
  {Nersisyan}, \citenamefont {Newman}, \citenamefont {Ng}, \citenamefont
  {Nguyen}, \citenamefont {Nguyen}, \citenamefont {Niu}, \citenamefont
  {O'Brien}, \citenamefont {Omonije}, \citenamefont {Petukhov}, \citenamefont
  {Potter}, \citenamefont {Pryadko}, \citenamefont {Quintana}, \citenamefont
  {Rocque}, \citenamefont {Rubin}, \citenamefont {Saei}, \citenamefont {Sank},
  \citenamefont {Sankaragomathi}, \citenamefont {Satzinger}, \citenamefont
  {Schurkus}, \citenamefont {Schuster}, \citenamefont {Shearn}, \citenamefont
  {Shorter}, \citenamefont {Shutty}, \citenamefont {Shvarts}, \citenamefont
  {Skruzny}, \citenamefont {Smith}, \citenamefont {Somma}, \citenamefont
  {Sterling}, \citenamefont {Strain}, \citenamefont {Szalay}, \citenamefont
  {Torres}, \citenamefont {Vidal}, \citenamefont {Villalonga}, \citenamefont
  {Heidweiller}, \citenamefont {White}, \citenamefont {Woo}, \citenamefont
  {Xing}, \citenamefont {Yao}, \citenamefont {Yeh}, \citenamefont {Yoo},
  \citenamefont {Young}, \citenamefont {Zalcman}, \citenamefont {Zhang},
  \citenamefont {Zhu}, \citenamefont {Zobrist}, \citenamefont {Neven},
  \citenamefont {Boixo}, \citenamefont {Megrant}, \citenamefont {Kelly},
  \citenamefont {Chen}, \citenamefont {Smelyanskiy}, \citenamefont {Kim},
  \citenamefont {Aleiner},\ and\ \citenamefont
  {Roushan}}]{google_non_abelian_2022}%
  \BibitemOpen
  \bibfield  {author} {\bibinfo {author} {\bibfnamefont {T.~I.}\ \bibnamefont
  {Andersen}}, \bibinfo {author} {\bibfnamefont {Y.~D.}\ \bibnamefont
  {Lensky}}, \bibinfo {author} {\bibfnamefont {K.}~\bibnamefont {Kechedzhi}},
  \bibinfo {author} {\bibfnamefont {I.}~\bibnamefont {Drozdov}}, \bibinfo
  {author} {\bibfnamefont {A.}~\bibnamefont {Bengtsson}}, \bibinfo {author}
  {\bibfnamefont {S.}~\bibnamefont {Hong}}, \bibinfo {author} {\bibfnamefont
  {A.}~\bibnamefont {Morvan}}, \bibinfo {author} {\bibfnamefont
  {X.}~\bibnamefont {Mi}}, \bibinfo {author} {\bibfnamefont {A.}~\bibnamefont
  {Opremcak}}, \bibinfo {author} {\bibfnamefont {R.}~\bibnamefont {Acharya}},
  \bibinfo {author} {\bibfnamefont {R.}~\bibnamefont {Allen}}, \bibinfo
  {author} {\bibfnamefont {M.}~\bibnamefont {Ansmann}}, \bibinfo {author}
  {\bibfnamefont {F.}~\bibnamefont {Arute}}, \bibinfo {author} {\bibfnamefont
  {K.}~\bibnamefont {Arya}}, \bibinfo {author} {\bibfnamefont {A.}~\bibnamefont
  {Asfaw}}, \bibinfo {author} {\bibfnamefont {J.}~\bibnamefont {Atalaya}},
  \bibinfo {author} {\bibfnamefont {R.}~\bibnamefont {Babbush}}, \bibinfo
  {author} {\bibfnamefont {D.}~\bibnamefont {Bacon}}, \bibinfo {author}
  {\bibfnamefont {J.~C.}\ \bibnamefont {Bardin}}, \bibinfo {author}
  {\bibfnamefont {G.}~\bibnamefont {Bortoli}}, \bibinfo {author} {\bibfnamefont
  {A.}~\bibnamefont {Bourassa}}, \bibinfo {author} {\bibfnamefont
  {J.}~\bibnamefont {Bovaird}}, \bibinfo {author} {\bibfnamefont
  {L.}~\bibnamefont {Brill}}, \bibinfo {author} {\bibfnamefont
  {M.}~\bibnamefont {Broughton}}, \bibinfo {author} {\bibfnamefont {B.~B.}\
  \bibnamefont {Buckley}}, \bibinfo {author} {\bibfnamefont {D.~A.}\
  \bibnamefont {Buell}}, \bibinfo {author} {\bibfnamefont {T.}~\bibnamefont
  {Burger}}, \bibinfo {author} {\bibfnamefont {B.}~\bibnamefont {Burkett}},
  \bibinfo {author} {\bibfnamefont {N.}~\bibnamefont {Bushnell}}, \bibinfo
  {author} {\bibfnamefont {Z.}~\bibnamefont {Chen}}, \bibinfo {author}
  {\bibfnamefont {B.}~\bibnamefont {Chiaro}}, \bibinfo {author} {\bibfnamefont
  {D.}~\bibnamefont {Chik}}, \bibinfo {author} {\bibfnamefont {C.}~\bibnamefont
  {Chou}}, \bibinfo {author} {\bibfnamefont {J.}~\bibnamefont {Cogan}},
  \bibinfo {author} {\bibfnamefont {R.}~\bibnamefont {Collins}}, \bibinfo
  {author} {\bibfnamefont {P.}~\bibnamefont {Conner}}, \bibinfo {author}
  {\bibfnamefont {W.}~\bibnamefont {Courtney}}, \bibinfo {author}
  {\bibfnamefont {A.~L.}\ \bibnamefont {Crook}}, \bibinfo {author}
  {\bibfnamefont {B.}~\bibnamefont {Curtin}}, \bibinfo {author} {\bibfnamefont
  {D.~M.}\ \bibnamefont {Debroy}}, \bibinfo {author} {\bibfnamefont {A.~D.~T.}\
  \bibnamefont {Barba}}, \bibinfo {author} {\bibfnamefont {S.}~\bibnamefont
  {Demura}}, \bibinfo {author} {\bibfnamefont {A.}~\bibnamefont {Dunsworth}},
  \bibinfo {author} {\bibfnamefont {D.}~\bibnamefont {Eppens}}, \bibinfo
  {author} {\bibfnamefont {C.}~\bibnamefont {Erickson}}, \bibinfo {author}
  {\bibfnamefont {L.}~\bibnamefont {Faoro}}, \bibinfo {author} {\bibfnamefont
  {E.}~\bibnamefont {Farhi}}, \bibinfo {author} {\bibfnamefont
  {R.}~\bibnamefont {Fatemi}}, \bibinfo {author} {\bibfnamefont {V.~S.}\
  \bibnamefont {Ferreira}}, \bibinfo {author} {\bibfnamefont {L.~F.}\
  \bibnamefont {Burgos}}, \bibinfo {author} {\bibfnamefont {E.}~\bibnamefont
  {Forati}}, \bibinfo {author} {\bibfnamefont {A.~G.}\ \bibnamefont {Fowler}},
  \bibinfo {author} {\bibfnamefont {B.}~\bibnamefont {Foxen}}, \bibinfo
  {author} {\bibfnamefont {W.}~\bibnamefont {Giang}}, \bibinfo {author}
  {\bibfnamefont {C.}~\bibnamefont {Gidney}}, \bibinfo {author} {\bibfnamefont
  {D.}~\bibnamefont {Gilboa}}, \bibinfo {author} {\bibfnamefont
  {M.}~\bibnamefont {Giustina}}, \bibinfo {author} {\bibfnamefont
  {R.}~\bibnamefont {Gosula}}, \bibinfo {author} {\bibfnamefont {A.~G.}\
  \bibnamefont {Dau}}, \bibinfo {author} {\bibfnamefont {J.~A.}\ \bibnamefont
  {Gross}}, \bibinfo {author} {\bibfnamefont {S.}~\bibnamefont {Habegger}},
  \bibinfo {author} {\bibfnamefont {M.~C.}\ \bibnamefont {Hamilton}}, \bibinfo
  {author} {\bibfnamefont {M.}~\bibnamefont {Hansen}}, \bibinfo {author}
  {\bibfnamefont {M.~P.}\ \bibnamefont {Harrigan}}, \bibinfo {author}
  {\bibfnamefont {S.~D.}\ \bibnamefont {Harrington}}, \bibinfo {author}
  {\bibfnamefont {P.}~\bibnamefont {Heu}}, \bibinfo {author} {\bibfnamefont
  {J.}~\bibnamefont {Hilton}}, \bibinfo {author} {\bibfnamefont {M.~R.}\
  \bibnamefont {Hoffmann}}, \bibinfo {author} {\bibfnamefont {T.}~\bibnamefont
  {Huang}}, \bibinfo {author} {\bibfnamefont {A.}~\bibnamefont {Huff}},
  \bibinfo {author} {\bibfnamefont {W.~J.}\ \bibnamefont {Huggins}}, \bibinfo
  {author} {\bibfnamefont {L.~B.}\ \bibnamefont {Ioffe}}, \bibinfo {author}
  {\bibfnamefont {S.~V.}\ \bibnamefont {Isakov}}, \bibinfo {author}
  {\bibfnamefont {J.}~\bibnamefont {Iveland}}, \bibinfo {author} {\bibfnamefont
  {E.}~\bibnamefont {Jeffrey}}, \bibinfo {author} {\bibfnamefont
  {Z.}~\bibnamefont {Jiang}}, \bibinfo {author} {\bibfnamefont
  {C.}~\bibnamefont {Jones}}, \bibinfo {author} {\bibfnamefont
  {P.}~\bibnamefont {Juhas}}, \bibinfo {author} {\bibfnamefont
  {D.}~\bibnamefont {Kafri}}, \bibinfo {author} {\bibfnamefont
  {T.}~\bibnamefont {Khattar}}, \bibinfo {author} {\bibfnamefont
  {M.}~\bibnamefont {Khezri}}, \bibinfo {author} {\bibfnamefont
  {M.}~\bibnamefont {Kieferová}}, \bibinfo {author} {\bibfnamefont
  {S.}~\bibnamefont {Kim}}, \bibinfo {author} {\bibfnamefont {A.}~\bibnamefont
  {Kitaev}}, \bibinfo {author} {\bibfnamefont {P.~V.}\ \bibnamefont {Klimov}},
  \bibinfo {author} {\bibfnamefont {A.~R.}\ \bibnamefont {Klots}}, \bibinfo
  {author} {\bibfnamefont {A.~N.}\ \bibnamefont {Korotkov}}, \bibinfo {author}
  {\bibfnamefont {F.}~\bibnamefont {Kostritsa}}, \bibinfo {author}
  {\bibfnamefont {J.~M.}\ \bibnamefont {Kreikebaum}}, \bibinfo {author}
  {\bibfnamefont {D.}~\bibnamefont {Landhuis}}, \bibinfo {author}
  {\bibfnamefont {P.}~\bibnamefont {Laptev}}, \bibinfo {author} {\bibfnamefont
  {K.-M.}\ \bibnamefont {Lau}}, \bibinfo {author} {\bibfnamefont
  {L.}~\bibnamefont {Laws}}, \bibinfo {author} {\bibfnamefont {J.}~\bibnamefont
  {Lee}}, \bibinfo {author} {\bibfnamefont {K.}~\bibnamefont {Lee}}, \bibinfo
  {author} {\bibfnamefont {B.~J.}\ \bibnamefont {Lester}}, \bibinfo {author}
  {\bibfnamefont {A.}~\bibnamefont {Lill}}, \bibinfo {author} {\bibfnamefont
  {W.}~\bibnamefont {Liu}}, \bibinfo {author} {\bibfnamefont {A.}~\bibnamefont
  {Locharla}}, \bibinfo {author} {\bibfnamefont {E.}~\bibnamefont {Lucero}},
  \bibinfo {author} {\bibfnamefont {F.~D.}\ \bibnamefont {Malone}}, \bibinfo
  {author} {\bibfnamefont {O.}~\bibnamefont {Martin}}, \bibinfo {author}
  {\bibfnamefont {J.~R.}\ \bibnamefont {McClean}}, \bibinfo {author}
  {\bibfnamefont {T.}~\bibnamefont {McCourt}}, \bibinfo {author} {\bibfnamefont
  {M.}~\bibnamefont {McEwen}}, \bibinfo {author} {\bibfnamefont {K.~C.}\
  \bibnamefont {Miao}}, \bibinfo {author} {\bibfnamefont {A.}~\bibnamefont
  {Mieszala}}, \bibinfo {author} {\bibfnamefont {M.}~\bibnamefont {Mohseni}},
  \bibinfo {author} {\bibfnamefont {S.}~\bibnamefont {Montazeri}}, \bibinfo
  {author} {\bibfnamefont {E.}~\bibnamefont {Mount}}, \bibinfo {author}
  {\bibfnamefont {R.}~\bibnamefont {Movassagh}}, \bibinfo {author}
  {\bibfnamefont {W.}~\bibnamefont {Mruczkiewicz}}, \bibinfo {author}
  {\bibfnamefont {O.}~\bibnamefont {Naaman}}, \bibinfo {author} {\bibfnamefont
  {M.}~\bibnamefont {Neeley}}, \bibinfo {author} {\bibfnamefont
  {C.}~\bibnamefont {Neill}}, \bibinfo {author} {\bibfnamefont
  {A.}~\bibnamefont {Nersisyan}}, \bibinfo {author} {\bibfnamefont
  {M.}~\bibnamefont {Newman}}, \bibinfo {author} {\bibfnamefont {J.~H.}\
  \bibnamefont {Ng}}, \bibinfo {author} {\bibfnamefont {A.}~\bibnamefont
  {Nguyen}}, \bibinfo {author} {\bibfnamefont {M.}~\bibnamefont {Nguyen}},
  \bibinfo {author} {\bibfnamefont {M.~Y.}\ \bibnamefont {Niu}}, \bibinfo
  {author} {\bibfnamefont {T.~E.}\ \bibnamefont {O'Brien}}, \bibinfo {author}
  {\bibfnamefont {S.}~\bibnamefont {Omonije}}, \bibinfo {author} {\bibfnamefont
  {A.}~\bibnamefont {Petukhov}}, \bibinfo {author} {\bibfnamefont
  {R.}~\bibnamefont {Potter}}, \bibinfo {author} {\bibfnamefont {L.~P.}\
  \bibnamefont {Pryadko}}, \bibinfo {author} {\bibfnamefont {C.}~\bibnamefont
  {Quintana}}, \bibinfo {author} {\bibfnamefont {C.}~\bibnamefont {Rocque}},
  \bibinfo {author} {\bibfnamefont {N.~C.}\ \bibnamefont {Rubin}}, \bibinfo
  {author} {\bibfnamefont {N.}~\bibnamefont {Saei}}, \bibinfo {author}
  {\bibfnamefont {D.}~\bibnamefont {Sank}}, \bibinfo {author} {\bibfnamefont
  {K.}~\bibnamefont {Sankaragomathi}}, \bibinfo {author} {\bibfnamefont
  {K.~J.}\ \bibnamefont {Satzinger}}, \bibinfo {author} {\bibfnamefont {H.~F.}\
  \bibnamefont {Schurkus}}, \bibinfo {author} {\bibfnamefont {C.}~\bibnamefont
  {Schuster}}, \bibinfo {author} {\bibfnamefont {M.~J.}\ \bibnamefont
  {Shearn}}, \bibinfo {author} {\bibfnamefont {A.}~\bibnamefont {Shorter}},
  \bibinfo {author} {\bibfnamefont {N.}~\bibnamefont {Shutty}}, \bibinfo
  {author} {\bibfnamefont {V.}~\bibnamefont {Shvarts}}, \bibinfo {author}
  {\bibfnamefont {J.}~\bibnamefont {Skruzny}}, \bibinfo {author} {\bibfnamefont
  {W.~C.}\ \bibnamefont {Smith}}, \bibinfo {author} {\bibfnamefont
  {R.}~\bibnamefont {Somma}}, \bibinfo {author} {\bibfnamefont
  {G.}~\bibnamefont {Sterling}}, \bibinfo {author} {\bibfnamefont
  {D.}~\bibnamefont {Strain}}, \bibinfo {author} {\bibfnamefont
  {M.}~\bibnamefont {Szalay}}, \bibinfo {author} {\bibfnamefont
  {A.}~\bibnamefont {Torres}}, \bibinfo {author} {\bibfnamefont
  {G.}~\bibnamefont {Vidal}}, \bibinfo {author} {\bibfnamefont
  {B.}~\bibnamefont {Villalonga}}, \bibinfo {author} {\bibfnamefont {C.~V.}\
  \bibnamefont {Heidweiller}}, \bibinfo {author} {\bibfnamefont
  {T.}~\bibnamefont {White}}, \bibinfo {author} {\bibfnamefont {B.~W.~K.}\
  \bibnamefont {Woo}}, \bibinfo {author} {\bibfnamefont {C.}~\bibnamefont
  {Xing}}, \bibinfo {author} {\bibfnamefont {Z.~J.}\ \bibnamefont {Yao}},
  \bibinfo {author} {\bibfnamefont {P.}~\bibnamefont {Yeh}}, \bibinfo {author}
  {\bibfnamefont {J.}~\bibnamefont {Yoo}}, \bibinfo {author} {\bibfnamefont
  {G.}~\bibnamefont {Young}}, \bibinfo {author} {\bibfnamefont
  {A.}~\bibnamefont {Zalcman}}, \bibinfo {author} {\bibfnamefont
  {Y.}~\bibnamefont {Zhang}}, \bibinfo {author} {\bibfnamefont
  {N.}~\bibnamefont {Zhu}}, \bibinfo {author} {\bibfnamefont {N.}~\bibnamefont
  {Zobrist}}, \bibinfo {author} {\bibfnamefont {H.}~\bibnamefont {Neven}},
  \bibinfo {author} {\bibfnamefont {S.}~\bibnamefont {Boixo}}, \bibinfo
  {author} {\bibfnamefont {A.}~\bibnamefont {Megrant}}, \bibinfo {author}
  {\bibfnamefont {J.}~\bibnamefont {Kelly}}, \bibinfo {author} {\bibfnamefont
  {Y.}~\bibnamefont {Chen}}, \bibinfo {author} {\bibfnamefont {V.}~\bibnamefont
  {Smelyanskiy}}, \bibinfo {author} {\bibfnamefont {E.-A.}\ \bibnamefont
  {Kim}}, \bibinfo {author} {\bibfnamefont {I.}~\bibnamefont {Aleiner}},\ and\
  \bibinfo {author} {\bibfnamefont {P.}~\bibnamefont {Roushan}},\ }\href@noop
  {} {\bibinfo {title} {Observation of non-{A}belian exchange statistics on a
  superconducting processor}} (\bibinfo {year} {2022}),\ \Eprint
  {https://arxiv.org/abs/2210.10255} {arXiv:2210.10255 [quant-ph]} \BibitemShut
  {NoStop}%
\bibitem [{\citenamefont {Haldane}(1991)}]{Haldane_1991}%
  \BibitemOpen
  \bibfield  {author} {\bibinfo {author} {\bibfnamefont {F.~D.~M.}\
  \bibnamefont {Haldane}},\ }\href {https://doi.org/10.1103/PhysRevLett.67.937}
  {\bibfield  {journal} {\bibinfo  {journal} {Phys. Rev. Lett.}\ }\textbf
  {\bibinfo {volume} {67}},\ \bibinfo {pages} {937} (\bibinfo {year}
  {1991})}\BibitemShut {NoStop}%
\bibitem [{\citenamefont {Greiter}(2009)}]{Greiter_2009}%
  \BibitemOpen
  \bibfield  {author} {\bibinfo {author} {\bibfnamefont {M.}~\bibnamefont
  {Greiter}},\ }\href {https://doi.org/10.1103/PhysRevB.79.064409} {\bibfield
  {journal} {\bibinfo  {journal} {Phys. Rev. B}\ }\textbf {\bibinfo {volume}
  {79}},\ \bibinfo {pages} {064409} (\bibinfo {year} {2009})}\BibitemShut
  {NoStop}%
\bibitem [{\citenamefont {Mourigal}\ \emph {et~al.}(2013)\citenamefont
  {Mourigal}, \citenamefont {Enderle}, \citenamefont {Klöpperpieper},
  \citenamefont {Caux}, \citenamefont {Stunault},\ and\ \citenamefont
  {R{\o}nnow}}]{Mourigal_2013}%
  \BibitemOpen
  \bibfield  {author} {\bibinfo {author} {\bibfnamefont {M.}~\bibnamefont
  {Mourigal}}, \bibinfo {author} {\bibfnamefont {M.}~\bibnamefont {Enderle}},
  \bibinfo {author} {\bibfnamefont {A.}~\bibnamefont {Klöpperpieper}},
  \bibinfo {author} {\bibfnamefont {J.-S.}\ \bibnamefont {Caux}}, \bibinfo
  {author} {\bibfnamefont {A.}~\bibnamefont {Stunault}},\ and\ \bibinfo
  {author} {\bibfnamefont {H.~M.}\ \bibnamefont {R{\o}nnow}},\ }\href
  {https://doi.org/10.1038/nphys2652} {\bibfield  {journal} {\bibinfo
  {journal} {Nat. Phys.}\ }\textbf {\bibinfo {volume} {9}},\ \bibinfo {pages}
  {435} (\bibinfo {year} {2013})}\BibitemShut {NoStop}%
\bibitem [{\citenamefont {Kundu}(1999)}]{1999_Kundu}%
  \BibitemOpen
  \bibfield  {author} {\bibinfo {author} {\bibfnamefont {A.}~\bibnamefont
  {Kundu}},\ }\href {https://doi.org/10.1103/PhysRevLett.83.1275} {\bibfield
  {journal} {\bibinfo  {journal} {Phys. Rev. Lett.}\ }\textbf {\bibinfo
  {volume} {83}},\ \bibinfo {pages} {1275} (\bibinfo {year}
  {1999})}\BibitemShut {NoStop}%
\bibitem [{\citenamefont {Harshman}\ and\ \citenamefont
  {Knapp}(2020)}]{Harshman_2020}%
  \BibitemOpen
  \bibfield  {author} {\bibinfo {author} {\bibfnamefont {N.}~\bibnamefont
  {Harshman}}\ and\ \bibinfo {author} {\bibfnamefont {A.}~\bibnamefont
  {Knapp}},\ }\href {https://doi.org/10.1016/j.aop.2019.168003} {\bibfield
  {journal} {\bibinfo  {journal} {Ann. Phys. (NY)}\ }\textbf {\bibinfo {volume}
  {412}},\ \bibinfo {pages} {168003} (\bibinfo {year} {2020})}\BibitemShut
  {NoStop}%
\bibitem [{\citenamefont {Bonkhoff}\ \emph {et~al.}(2021)\citenamefont
  {Bonkhoff}, \citenamefont {J\"agering}, \citenamefont {Eggert}, \citenamefont
  {Pelster}, \citenamefont {Thorwart},\ and\ \citenamefont
  {Posske}}]{2021_Bonkhoff}%
  \BibitemOpen
  \bibfield  {author} {\bibinfo {author} {\bibfnamefont {M.}~\bibnamefont
  {Bonkhoff}}, \bibinfo {author} {\bibfnamefont {K.}~\bibnamefont
  {J\"agering}}, \bibinfo {author} {\bibfnamefont {S.}~\bibnamefont {Eggert}},
  \bibinfo {author} {\bibfnamefont {A.}~\bibnamefont {Pelster}}, \bibinfo
  {author} {\bibfnamefont {M.}~\bibnamefont {Thorwart}},\ and\ \bibinfo
  {author} {\bibfnamefont {T.}~\bibnamefont {Posske}},\ }\href
  {https://doi.org/10.1103/PhysRevLett.126.163201} {\bibfield  {journal}
  {\bibinfo  {journal} {Phys. Rev. Lett.}\ }\textbf {\bibinfo {volume} {126}},\
  \bibinfo {pages} {163201} (\bibinfo {year} {2021})}\BibitemShut {NoStop}%
\bibitem [{\citenamefont {Keilmann}\ \emph {et~al.}(2011)\citenamefont
  {Keilmann}, \citenamefont {Lanzmich}, \citenamefont {McCulloch},\ and\
  \citenamefont {Roncaglia}}]{Keilmann_2011}%
  \BibitemOpen
  \bibfield  {author} {\bibinfo {author} {\bibfnamefont {T.}~\bibnamefont
  {Keilmann}}, \bibinfo {author} {\bibfnamefont {S.}~\bibnamefont {Lanzmich}},
  \bibinfo {author} {\bibfnamefont {I.}~\bibnamefont {McCulloch}},\ and\
  \bibinfo {author} {\bibfnamefont {M.}~\bibnamefont {Roncaglia}},\ }\href
  {https://doi.org/10.1038/ncomms1353} {\bibfield  {journal} {\bibinfo
  {journal} {Nat. Commun.}\ }\textbf {\bibinfo {volume} {2}},\ \bibinfo {pages}
  {361} (\bibinfo {year} {2011})}\BibitemShut {NoStop}%
\bibitem [{\citenamefont {Hao}\ \emph {et~al.}(2008)\citenamefont {Hao},
  \citenamefont {Zhang},\ and\ \citenamefont {Chen}}]{Hao_2008}%
  \BibitemOpen
  \bibfield  {author} {\bibinfo {author} {\bibfnamefont {Y.}~\bibnamefont
  {Hao}}, \bibinfo {author} {\bibfnamefont {Y.}~\bibnamefont {Zhang}},\ and\
  \bibinfo {author} {\bibfnamefont {S.}~\bibnamefont {Chen}},\ }\href
  {https://doi.org/10.1103/PhysRevA.78.023631} {\bibfield  {journal} {\bibinfo
  {journal} {Phys. Rev. A}\ }\textbf {\bibinfo {volume} {78}},\ \bibinfo
  {pages} {023631} (\bibinfo {year} {2008})}\BibitemShut {NoStop}%
\bibitem [{\citenamefont {Hao}\ \emph {et~al.}(2009)\citenamefont {Hao},
  \citenamefont {Zhang},\ and\ \citenamefont {Chen}}]{Hao_2009}%
  \BibitemOpen
  \bibfield  {author} {\bibinfo {author} {\bibfnamefont {Y.}~\bibnamefont
  {Hao}}, \bibinfo {author} {\bibfnamefont {Y.}~\bibnamefont {Zhang}},\ and\
  \bibinfo {author} {\bibfnamefont {S.}~\bibnamefont {Chen}},\ }\href
  {https://doi.org/10.1103/PhysRevA.79.043633} {\bibfield  {journal} {\bibinfo
  {journal} {Phys. Rev. A}\ }\textbf {\bibinfo {volume} {79}},\ \bibinfo
  {pages} {043633} (\bibinfo {year} {2009})}\BibitemShut {NoStop}%
\bibitem [{\citenamefont {Tang}\ \emph {et~al.}(2015)\citenamefont {Tang},
  \citenamefont {Eggert},\ and\ \citenamefont {Pelster}}]{Tang_2015}%
  \BibitemOpen
  \bibfield  {author} {\bibinfo {author} {\bibfnamefont {G.}~\bibnamefont
  {Tang}}, \bibinfo {author} {\bibfnamefont {S.}~\bibnamefont {Eggert}},\ and\
  \bibinfo {author} {\bibfnamefont {A.}~\bibnamefont {Pelster}},\ }\href
  {https://doi.org/10.1088/1367-2630/17/12/123016} {\bibfield  {journal}
  {\bibinfo  {journal} {New J. Phys.}\ }\textbf {\bibinfo {volume} {17}},\
  \bibinfo {pages} {123016} (\bibinfo {year} {2015})}\BibitemShut {NoStop}%
\bibitem [{\citenamefont {Str\"ater}\ \emph {et~al.}(2016)\citenamefont
  {Str\"ater}, \citenamefont {Srivastava},\ and\ \citenamefont
  {Eckardt}}]{2016_Strater}%
  \BibitemOpen
  \bibfield  {author} {\bibinfo {author} {\bibfnamefont {C.}~\bibnamefont
  {Str\"ater}}, \bibinfo {author} {\bibfnamefont {S.~C.~L.}\ \bibnamefont
  {Srivastava}},\ and\ \bibinfo {author} {\bibfnamefont {A.}~\bibnamefont
  {Eckardt}},\ }\href {https://doi.org/10.1103/PhysRevLett.117.205303}
  {\bibfield  {journal} {\bibinfo  {journal} {Phys. Rev. Lett.}\ }\textbf
  {\bibinfo {volume} {117}},\ \bibinfo {pages} {205303} (\bibinfo {year}
  {2016})}\BibitemShut {NoStop}%
\bibitem [{\citenamefont {Yuan}\ \emph {et~al.}(2017)\citenamefont {Yuan},
  \citenamefont {Xiao}, \citenamefont {Xu},\ and\ \citenamefont
  {Fan}}]{Yuan_2017}%
  \BibitemOpen
  \bibfield  {author} {\bibinfo {author} {\bibfnamefont {L.}~\bibnamefont
  {Yuan}}, \bibinfo {author} {\bibfnamefont {M.}~\bibnamefont {Xiao}}, \bibinfo
  {author} {\bibfnamefont {S.}~\bibnamefont {Xu}},\ and\ \bibinfo {author}
  {\bibfnamefont {S.}~\bibnamefont {Fan}},\ }\href
  {https://doi.org/10.1103/PhysRevA.96.043864} {\bibfield  {journal} {\bibinfo
  {journal} {Phys. Rev. A}\ }\textbf {\bibinfo {volume} {96}},\ \bibinfo
  {pages} {043864} (\bibinfo {year} {2017})}\BibitemShut {NoStop}%
\bibitem [{\citenamefont {Greschner}\ and\ \citenamefont
  {Santos}(2015)}]{2015_Greschner}%
  \BibitemOpen
  \bibfield  {author} {\bibinfo {author} {\bibfnamefont {S.}~\bibnamefont
  {Greschner}}\ and\ \bibinfo {author} {\bibfnamefont {L.}~\bibnamefont
  {Santos}},\ }\href {https://doi.org/10.1103/PhysRevLett.115.053002}
  {\bibfield  {journal} {\bibinfo  {journal} {Phys. Rev. Lett.}\ }\textbf
  {\bibinfo {volume} {115}},\ \bibinfo {pages} {053002} (\bibinfo {year}
  {2015})}\BibitemShut {NoStop}%
\bibitem [{\citenamefont {Zhang}\ \emph {et~al.}(2017)\citenamefont {Zhang},
  \citenamefont {Greschner}, \citenamefont {Fan}, \citenamefont {Scott},\ and\
  \citenamefont {Zhang}}]{2017_Zhang}%
  \BibitemOpen
  \bibfield  {author} {\bibinfo {author} {\bibfnamefont {W.}~\bibnamefont
  {Zhang}}, \bibinfo {author} {\bibfnamefont {S.}~\bibnamefont {Greschner}},
  \bibinfo {author} {\bibfnamefont {E.}~\bibnamefont {Fan}}, \bibinfo {author}
  {\bibfnamefont {T.~C.}\ \bibnamefont {Scott}},\ and\ \bibinfo {author}
  {\bibfnamefont {Y.}~\bibnamefont {Zhang}},\ }\href
  {https://doi.org/10.1103/PhysRevA.95.053614} {\bibfield  {journal} {\bibinfo
  {journal} {Phys. Rev. A}\ }\textbf {\bibinfo {volume} {95}},\ \bibinfo
  {pages} {053614} (\bibinfo {year} {2017})}\BibitemShut {NoStop}%
\bibitem [{\citenamefont {Sansoni}\ \emph {et~al.}(2012)\citenamefont
  {Sansoni}, \citenamefont {Sciarrino}, \citenamefont {Vallone}, \citenamefont
  {Mataloni}, \citenamefont {Crespi}, \citenamefont {Ramponi},\ and\
  \citenamefont {Osellame}}]{Sansoni_2012}%
  \BibitemOpen
  \bibfield  {author} {\bibinfo {author} {\bibfnamefont {L.}~\bibnamefont
  {Sansoni}}, \bibinfo {author} {\bibfnamefont {F.}~\bibnamefont {Sciarrino}},
  \bibinfo {author} {\bibfnamefont {G.}~\bibnamefont {Vallone}}, \bibinfo
  {author} {\bibfnamefont {P.}~\bibnamefont {Mataloni}}, \bibinfo {author}
  {\bibfnamefont {A.}~\bibnamefont {Crespi}}, \bibinfo {author} {\bibfnamefont
  {R.}~\bibnamefont {Ramponi}},\ and\ \bibinfo {author} {\bibfnamefont
  {R.}~\bibnamefont {Osellame}},\ }\href
  {https://doi.org/10.1103/PhysRevLett.108.010502} {\bibfield  {journal}
  {\bibinfo  {journal} {Phys. Rev. Lett.}\ }\textbf {\bibinfo {volume} {108}},\
  \bibinfo {pages} {010502} (\bibinfo {year} {2012})}\BibitemShut {NoStop}%
\bibitem [{\citenamefont {Matthews}\ \emph {et~al.}(2013)\citenamefont
  {Matthews}, \citenamefont {Poulios}, \citenamefont {Meinecke}, \citenamefont
  {Politi}, \citenamefont {Peruzzo}, \citenamefont {Ismail}, \citenamefont
  {Wörhoff}, \citenamefont {Thompson},\ and\ \citenamefont
  {O{\textquotesingle}Brien}}]{Matthews_2013}%
  \BibitemOpen
  \bibfield  {author} {\bibinfo {author} {\bibfnamefont {J.~C.~F.}\
  \bibnamefont {Matthews}}, \bibinfo {author} {\bibfnamefont {K.}~\bibnamefont
  {Poulios}}, \bibinfo {author} {\bibfnamefont {J.~D.~A.}\ \bibnamefont
  {Meinecke}}, \bibinfo {author} {\bibfnamefont {A.}~\bibnamefont {Politi}},
  \bibinfo {author} {\bibfnamefont {A.}~\bibnamefont {Peruzzo}}, \bibinfo
  {author} {\bibfnamefont {N.}~\bibnamefont {Ismail}}, \bibinfo {author}
  {\bibfnamefont {K.}~\bibnamefont {Wörhoff}}, \bibinfo {author}
  {\bibfnamefont {M.~G.}\ \bibnamefont {Thompson}},\ and\ \bibinfo {author}
  {\bibfnamefont {J.~L.}\ \bibnamefont {O{\textquotesingle}Brien}},\ }\href
  {https://doi.org/10.1038/srep01539} {\bibfield  {journal} {\bibinfo
  {journal} {Sci. Rep.}\ }\textbf {\bibinfo {volume} {3}},\ \bibinfo {pages}
  {1539} (\bibinfo {year} {2013})}\BibitemShut {NoStop}%
\bibitem [{\citenamefont {Zhang}\ \emph {et~al.}(2022)\citenamefont {Zhang},
  \citenamefont {Yuan}, \citenamefont {Wang}, \citenamefont {Di}, \citenamefont
  {Sun}, \citenamefont {Zheng}, \citenamefont {Sun},\ and\ \citenamefont
  {Zhang}}]{zhang_observation_2022}%
  \BibitemOpen
  \bibfield  {author} {\bibinfo {author} {\bibfnamefont {W.}~\bibnamefont
  {Zhang}}, \bibinfo {author} {\bibfnamefont {H.}~\bibnamefont {Yuan}},
  \bibinfo {author} {\bibfnamefont {H.}~\bibnamefont {Wang}}, \bibinfo {author}
  {\bibfnamefont {F.}~\bibnamefont {Di}}, \bibinfo {author} {\bibfnamefont
  {N.}~\bibnamefont {Sun}}, \bibinfo {author} {\bibfnamefont {X.}~\bibnamefont
  {Zheng}}, \bibinfo {author} {\bibfnamefont {H.}~\bibnamefont {Sun}},\ and\
  \bibinfo {author} {\bibfnamefont {X.}~\bibnamefont {Zhang}},\ }\href
  {https://doi.org/10.1038/s41467-022-29895-0} {\bibfield  {journal} {\bibinfo
  {journal} {Nat. Commun.}\ }\textbf {\bibinfo {volume} {13}},\ \bibinfo
  {pages} {2392} (\bibinfo {year} {2022})}\BibitemShut {NoStop}%
\bibitem [{\citenamefont {Bakr}\ \emph {et~al.}(2009)\citenamefont {Bakr},
  \citenamefont {Gillen}, \citenamefont {Peng}, \citenamefont {Fölling},\ and\
  \citenamefont {Greiner}}]{Bakr_2009}%
  \BibitemOpen
  \bibfield  {author} {\bibinfo {author} {\bibfnamefont {W.~S.}\ \bibnamefont
  {Bakr}}, \bibinfo {author} {\bibfnamefont {J.~I.}\ \bibnamefont {Gillen}},
  \bibinfo {author} {\bibfnamefont {A.}~\bibnamefont {Peng}}, \bibinfo {author}
  {\bibfnamefont {S.}~\bibnamefont {Fölling}},\ and\ \bibinfo {author}
  {\bibfnamefont {M.}~\bibnamefont {Greiner}},\ }\href
  {https://doi.org/10.1038/nature08482} {\bibfield  {journal} {\bibinfo
  {journal} {Nature}\ }\textbf {\bibinfo {volume} {462}},\ \bibinfo {pages}
  {74} (\bibinfo {year} {2009})}\BibitemShut {NoStop}%
\bibitem [{\citenamefont {Wang}\ \emph {et~al.}(2014)\citenamefont {Wang},
  \citenamefont {Wang},\ and\ \citenamefont {Zhang}}]{Wang_2014}%
  \BibitemOpen
  \bibfield  {author} {\bibinfo {author} {\bibfnamefont {L.}~\bibnamefont
  {Wang}}, \bibinfo {author} {\bibfnamefont {L.}~\bibnamefont {Wang}},\ and\
  \bibinfo {author} {\bibfnamefont {Y.}~\bibnamefont {Zhang}},\ }\href
  {https://doi.org/10.1103/PhysRevA.90.063618} {\bibfield  {journal} {\bibinfo
  {journal} {Phys. Rev. A}\ }\textbf {\bibinfo {volume} {90}},\ \bibinfo
  {pages} {063618} (\bibinfo {year} {2014})}\BibitemShut {NoStop}%
\bibitem [{\citenamefont {Preiss}\ \emph {et~al.}(2015)\citenamefont {Preiss},
  \citenamefont {Ma}, \citenamefont {Tai}, \citenamefont {Lukin}, \citenamefont
  {Rispoli}, \citenamefont {Zupancic}, \citenamefont {Lahini}, \citenamefont
  {Islam},\ and\ \citenamefont {Greiner}}]{Preiss_2015}%
  \BibitemOpen
  \bibfield  {author} {\bibinfo {author} {\bibfnamefont {P.~M.}\ \bibnamefont
  {Preiss}}, \bibinfo {author} {\bibfnamefont {R.}~\bibnamefont {Ma}}, \bibinfo
  {author} {\bibfnamefont {M.~E.}\ \bibnamefont {Tai}}, \bibinfo {author}
  {\bibfnamefont {A.}~\bibnamefont {Lukin}}, \bibinfo {author} {\bibfnamefont
  {M.}~\bibnamefont {Rispoli}}, \bibinfo {author} {\bibfnamefont
  {P.}~\bibnamefont {Zupancic}}, \bibinfo {author} {\bibfnamefont
  {Y.}~\bibnamefont {Lahini}}, \bibinfo {author} {\bibfnamefont
  {R.}~\bibnamefont {Islam}},\ and\ \bibinfo {author} {\bibfnamefont
  {M.}~\bibnamefont {Greiner}},\ }\href
  {https://doi.org/10.1126/science.1260364} {\bibfield  {journal} {\bibinfo
  {journal} {Science}\ }\textbf {\bibinfo {volume} {347}},\ \bibinfo {pages}
  {1229} (\bibinfo {year} {2015})}\BibitemShut {NoStop}%
\bibitem [{\citenamefont {Clark}\ \emph {et~al.}(2018)\citenamefont {Clark},
  \citenamefont {Anderson}, \citenamefont {Feng}, \citenamefont {Gaj},
  \citenamefont {Levin},\ and\ \citenamefont {Chin}}]{Clark_2018}%
  \BibitemOpen
  \bibfield  {author} {\bibinfo {author} {\bibfnamefont {L.~W.}\ \bibnamefont
  {Clark}}, \bibinfo {author} {\bibfnamefont {B.~M.}\ \bibnamefont {Anderson}},
  \bibinfo {author} {\bibfnamefont {L.}~\bibnamefont {Feng}}, \bibinfo {author}
  {\bibfnamefont {A.}~\bibnamefont {Gaj}}, \bibinfo {author} {\bibfnamefont
  {K.}~\bibnamefont {Levin}},\ and\ \bibinfo {author} {\bibfnamefont
  {C.}~\bibnamefont {Chin}},\ }\href
  {https://doi.org/10.1103/PhysRevLett.121.030402} {\bibfield  {journal}
  {\bibinfo  {journal} {Phys. Rev. Lett.}\ }\textbf {\bibinfo {volume} {121}},\
  \bibinfo {pages} {030402} (\bibinfo {year} {2018})}\BibitemShut {NoStop}%
\bibitem [{\citenamefont {Görg}\ \emph {et~al.}(2019)\citenamefont {Görg},
  \citenamefont {Sandholzer}, \citenamefont {Minguzzi}, \citenamefont
  {Desbuquois}, \citenamefont {Messer},\ and\ \citenamefont
  {Esslinger}}]{G_rg_2019}%
  \BibitemOpen
  \bibfield  {author} {\bibinfo {author} {\bibfnamefont {F.}~\bibnamefont
  {Görg}}, \bibinfo {author} {\bibfnamefont {K.}~\bibnamefont {Sandholzer}},
  \bibinfo {author} {\bibfnamefont {J.}~\bibnamefont {Minguzzi}}, \bibinfo
  {author} {\bibfnamefont {R.}~\bibnamefont {Desbuquois}}, \bibinfo {author}
  {\bibfnamefont {M.}~\bibnamefont {Messer}},\ and\ \bibinfo {author}
  {\bibfnamefont {T.}~\bibnamefont {Esslinger}},\ }\href
  {https://doi.org/10.1038/s41567-019-0615-4} {\bibfield  {journal} {\bibinfo
  {journal} {Nat. Phys.}\ }\textbf {\bibinfo {volume} {15}},\ \bibinfo {pages}
  {1161} (\bibinfo {year} {2019})}\BibitemShut {NoStop}%
\bibitem [{\citenamefont {Schweizer}\ \emph {et~al.}(2019)\citenamefont
  {Schweizer}, \citenamefont {Grusdt}, \citenamefont {Berngruber},
  \citenamefont {Barbiero}, \citenamefont {Demler}, \citenamefont {Goldman},
  \citenamefont {Bloch},\ and\ \citenamefont {Aidelsburger}}]{Schweizer_2019}%
  \BibitemOpen
  \bibfield  {author} {\bibinfo {author} {\bibfnamefont {C.}~\bibnamefont
  {Schweizer}}, \bibinfo {author} {\bibfnamefont {F.}~\bibnamefont {Grusdt}},
  \bibinfo {author} {\bibfnamefont {M.}~\bibnamefont {Berngruber}}, \bibinfo
  {author} {\bibfnamefont {L.}~\bibnamefont {Barbiero}}, \bibinfo {author}
  {\bibfnamefont {E.}~\bibnamefont {Demler}}, \bibinfo {author} {\bibfnamefont
  {N.}~\bibnamefont {Goldman}}, \bibinfo {author} {\bibfnamefont
  {I.}~\bibnamefont {Bloch}},\ and\ \bibinfo {author} {\bibfnamefont
  {M.}~\bibnamefont {Aidelsburger}},\ }\href
  {https://doi.org/10.1038/s41567-019-0649-7} {\bibfield  {journal} {\bibinfo
  {journal} {Nat. Phys.}\ }\textbf {\bibinfo {volume} {15}},\ \bibinfo {pages}
  {1168} (\bibinfo {year} {2019})}\BibitemShut {NoStop}%
\bibitem [{\citenamefont {Wilczek}(1990)}]{Wilczek_1990}%
  \BibitemOpen
  \bibfield  {author} {\bibinfo {author} {\bibfnamefont {F.}~\bibnamefont
  {Wilczek}},\ }\href@noop {} {\emph {\bibinfo {title} {Fractional Statistics
  and Anyon Superconductivity}}}\ (\bibinfo  {publisher} {World Scientific},\
  \bibinfo {year} {1990})\BibitemShut {NoStop}%
\bibitem [{any()}]{anyons_sm}%
  \BibitemOpen
  \href@noop {} {\bibinfo {title} {Materials and methods are available as
  supplementary materials.}}\BibitemShut {Stop}%
\bibitem [{\citenamefont {Cardarelli}\ \emph {et~al.}(2016)\citenamefont
  {Cardarelli}, \citenamefont {Greschner},\ and\ \citenamefont
  {Santos}}]{2016_Cardarelli}%
  \BibitemOpen
  \bibfield  {author} {\bibinfo {author} {\bibfnamefont {L.}~\bibnamefont
  {Cardarelli}}, \bibinfo {author} {\bibfnamefont {S.}~\bibnamefont
  {Greschner}},\ and\ \bibinfo {author} {\bibfnamefont {L.}~\bibnamefont
  {Santos}},\ }\href {https://doi.org/10.1103/PhysRevA.94.023615} {\bibfield
  {journal} {\bibinfo  {journal} {Phys. Rev. A}\ }\textbf {\bibinfo {volume}
  {94}},\ \bibinfo {pages} {023615} (\bibinfo {year} {2016})}\BibitemShut
  {NoStop}%
\bibitem [{\citenamefont {Zupancic}\ \emph {et~al.}(2016)\citenamefont
  {Zupancic}, \citenamefont {Preiss}, \citenamefont {Ma}, \citenamefont
  {Lukin}, \citenamefont {Tai}, \citenamefont {Rispoli}, \citenamefont
  {Islam},\ and\ \citenamefont {Greiner}}]{Zupancic_2016}%
  \BibitemOpen
  \bibfield  {author} {\bibinfo {author} {\bibfnamefont {P.}~\bibnamefont
  {Zupancic}}, \bibinfo {author} {\bibfnamefont {P.~M.}\ \bibnamefont
  {Preiss}}, \bibinfo {author} {\bibfnamefont {R.}~\bibnamefont {Ma}}, \bibinfo
  {author} {\bibfnamefont {A.}~\bibnamefont {Lukin}}, \bibinfo {author}
  {\bibfnamefont {M.~E.}\ \bibnamefont {Tai}}, \bibinfo {author} {\bibfnamefont
  {M.}~\bibnamefont {Rispoli}}, \bibinfo {author} {\bibfnamefont
  {R.}~\bibnamefont {Islam}},\ and\ \bibinfo {author} {\bibfnamefont
  {M.}~\bibnamefont {Greiner}},\ }\href {https://doi.org/10.1364/oe.24.013881}
  {\bibfield  {journal} {\bibinfo  {journal} {Opt. Express}\ }\textbf {\bibinfo
  {volume} {24}},\ \bibinfo {pages} {13881} (\bibinfo {year}
  {2016})}\BibitemShut {NoStop}%
\bibitem [{\citenamefont {Henny}\ \emph {et~al.}(1999)\citenamefont {Henny},
  \citenamefont {Oberholzer}, \citenamefont {Strunk}, \citenamefont {Heinzel},
  \citenamefont {Ensslin}, \citenamefont {Holland},\ and\ \citenamefont
  {Schönenberger}}]{Henny_1999}%
  \BibitemOpen
  \bibfield  {author} {\bibinfo {author} {\bibfnamefont {M.}~\bibnamefont
  {Henny}}, \bibinfo {author} {\bibfnamefont {S.}~\bibnamefont {Oberholzer}},
  \bibinfo {author} {\bibfnamefont {C.}~\bibnamefont {Strunk}}, \bibinfo
  {author} {\bibfnamefont {T.}~\bibnamefont {Heinzel}}, \bibinfo {author}
  {\bibfnamefont {K.}~\bibnamefont {Ensslin}}, \bibinfo {author} {\bibfnamefont
  {M.}~\bibnamefont {Holland}},\ and\ \bibinfo {author} {\bibfnamefont
  {C.}~\bibnamefont {Schönenberger}},\ }\href
  {https://doi.org/10.1126/science.284.5412.296} {\bibfield  {journal}
  {\bibinfo  {journal} {Science}\ }\textbf {\bibinfo {volume} {284}},\ \bibinfo
  {pages} {296} (\bibinfo {year} {1999})}\BibitemShut {NoStop}%
\bibitem [{\citenamefont {Jeltes}\ \emph {et~al.}(2007)\citenamefont {Jeltes},
  \citenamefont {McNamara}, \citenamefont {Hogervorst}, \citenamefont {Vassen},
  \citenamefont {Krachmalnicoff}, \citenamefont {Schellekens}, \citenamefont
  {Perrin}, \citenamefont {Chang}, \citenamefont {Boiron}, \citenamefont
  {Aspect},\ and\ \citenamefont {Westbrook}}]{Jeltes_2007}%
  \BibitemOpen
  \bibfield  {author} {\bibinfo {author} {\bibfnamefont {T.}~\bibnamefont
  {Jeltes}}, \bibinfo {author} {\bibfnamefont {J.~M.}\ \bibnamefont
  {McNamara}}, \bibinfo {author} {\bibfnamefont {W.}~\bibnamefont
  {Hogervorst}}, \bibinfo {author} {\bibfnamefont {W.}~\bibnamefont {Vassen}},
  \bibinfo {author} {\bibfnamefont {V.}~\bibnamefont {Krachmalnicoff}},
  \bibinfo {author} {\bibfnamefont {M.}~\bibnamefont {Schellekens}}, \bibinfo
  {author} {\bibfnamefont {A.}~\bibnamefont {Perrin}}, \bibinfo {author}
  {\bibfnamefont {H.}~\bibnamefont {Chang}}, \bibinfo {author} {\bibfnamefont
  {D.}~\bibnamefont {Boiron}}, \bibinfo {author} {\bibfnamefont
  {A.}~\bibnamefont {Aspect}},\ and\ \bibinfo {author} {\bibfnamefont {C.~I.}\
  \bibnamefont {Westbrook}},\ }\href {https://doi.org/10.1038/nature05513}
  {\bibfield  {journal} {\bibinfo  {journal} {Nature}\ }\textbf {\bibinfo
  {volume} {445}},\ \bibinfo {pages} {402} (\bibinfo {year}
  {2007})}\BibitemShut {NoStop}%
\bibitem [{\citenamefont {Peruzzo}\ \emph {et~al.}(2010)\citenamefont
  {Peruzzo}, \citenamefont {Lobino}, \citenamefont {Matthews}, \citenamefont
  {Matsuda}, \citenamefont {Politi}, \citenamefont {Poulios}, \citenamefont
  {Zhou}, \citenamefont {Lahini}, \citenamefont {Ismail}, \citenamefont
  {Wörhoff}, \citenamefont {Bromberg}, \citenamefont {Silberberg},
  \citenamefont {Thompson},\ and\ \citenamefont
  {O{\textquotesingle}Brien}}]{Peruzzo_2010}%
  \BibitemOpen
  \bibfield  {author} {\bibinfo {author} {\bibfnamefont {A.}~\bibnamefont
  {Peruzzo}}, \bibinfo {author} {\bibfnamefont {M.}~\bibnamefont {Lobino}},
  \bibinfo {author} {\bibfnamefont {J.~C.~F.}\ \bibnamefont {Matthews}},
  \bibinfo {author} {\bibfnamefont {N.}~\bibnamefont {Matsuda}}, \bibinfo
  {author} {\bibfnamefont {A.}~\bibnamefont {Politi}}, \bibinfo {author}
  {\bibfnamefont {K.}~\bibnamefont {Poulios}}, \bibinfo {author} {\bibfnamefont
  {X.-Q.}\ \bibnamefont {Zhou}}, \bibinfo {author} {\bibfnamefont
  {Y.}~\bibnamefont {Lahini}}, \bibinfo {author} {\bibfnamefont
  {N.}~\bibnamefont {Ismail}}, \bibinfo {author} {\bibfnamefont
  {K.}~\bibnamefont {Wörhoff}}, \bibinfo {author} {\bibfnamefont
  {Y.}~\bibnamefont {Bromberg}}, \bibinfo {author} {\bibfnamefont
  {Y.}~\bibnamefont {Silberberg}}, \bibinfo {author} {\bibfnamefont {M.~G.}\
  \bibnamefont {Thompson}},\ and\ \bibinfo {author} {\bibfnamefont {J.~L.}\
  \bibnamefont {O{\textquotesingle}Brien}},\ }\href
  {https://doi.org/10.1126/science.1193515} {\bibfield  {journal} {\bibinfo
  {journal} {Science}\ }\textbf {\bibinfo {volume} {329}},\ \bibinfo {pages}
  {1500} (\bibinfo {year} {2010})}\BibitemShut {NoStop}%
\bibitem [{\citenamefont {Winkler}\ \emph {et~al.}(2006)\citenamefont
  {Winkler}, \citenamefont {Thalhammer}, \citenamefont {Lang}, \citenamefont
  {Grimm}, \citenamefont {Hecker~Denschlag}, \citenamefont {Daley},
  \citenamefont {Kantian}, \citenamefont {B{\"u}chler},\ and\ \citenamefont
  {Zoller}}]{2006_Winkler}%
  \BibitemOpen
  \bibfield  {author} {\bibinfo {author} {\bibfnamefont {K.}~\bibnamefont
  {Winkler}}, \bibinfo {author} {\bibfnamefont {G.}~\bibnamefont {Thalhammer}},
  \bibinfo {author} {\bibfnamefont {F.}~\bibnamefont {Lang}}, \bibinfo {author}
  {\bibfnamefont {R.}~\bibnamefont {Grimm}}, \bibinfo {author} {\bibfnamefont
  {J.}~\bibnamefont {Hecker~Denschlag}}, \bibinfo {author} {\bibfnamefont
  {A.~J.}\ \bibnamefont {Daley}}, \bibinfo {author} {\bibfnamefont
  {A.}~\bibnamefont {Kantian}}, \bibinfo {author} {\bibfnamefont {H.~P.}\
  \bibnamefont {B{\"u}chler}},\ and\ \bibinfo {author} {\bibfnamefont
  {P.}~\bibnamefont {Zoller}},\ }\href {https://doi.org/10.1038/nature04918}
  {\bibfield  {journal} {\bibinfo  {journal} {Nature}\ }\textbf {\bibinfo
  {volume} {441}},\ \bibinfo {pages} {853} (\bibinfo {year}
  {2006})}\BibitemShut {NoStop}%
\bibitem [{\citenamefont {Fukuhara}\ \emph {et~al.}(2013)\citenamefont
  {Fukuhara}, \citenamefont {Schau{\ss}}, \citenamefont {Endres}, \citenamefont
  {Hild}, \citenamefont {Cheneau}, \citenamefont {Bloch},\ and\ \citenamefont
  {Gross}}]{2013_Fukuhara}%
  \BibitemOpen
  \bibfield  {author} {\bibinfo {author} {\bibfnamefont {T.}~\bibnamefont
  {Fukuhara}}, \bibinfo {author} {\bibfnamefont {P.}~\bibnamefont
  {Schau{\ss}}}, \bibinfo {author} {\bibfnamefont {M.}~\bibnamefont {Endres}},
  \bibinfo {author} {\bibfnamefont {S.}~\bibnamefont {Hild}}, \bibinfo {author}
  {\bibfnamefont {M.}~\bibnamefont {Cheneau}}, \bibinfo {author} {\bibfnamefont
  {I.}~\bibnamefont {Bloch}},\ and\ \bibinfo {author} {\bibfnamefont
  {C.}~\bibnamefont {Gross}},\ }\href {https://doi.org/10.1038/nature12541}
  {\bibfield  {journal} {\bibinfo  {journal} {Nature}\ }\textbf {\bibinfo
  {volume} {502}},\ \bibinfo {pages} {76} (\bibinfo {year} {2013})}\BibitemShut
  {NoStop}%
\bibitem [{\citenamefont {Kranzl}\ \emph {et~al.}(2022)\citenamefont {Kranzl},
  \citenamefont {Birnkammer}, \citenamefont {Joshi}, \citenamefont
  {Bastianello}, \citenamefont {Blatt}, \citenamefont {Knap},\ and\
  \citenamefont {Roos}}]{2022_Kranzl}%
  \BibitemOpen
  \bibfield  {author} {\bibinfo {author} {\bibfnamefont {F.}~\bibnamefont
  {Kranzl}}, \bibinfo {author} {\bibfnamefont {S.}~\bibnamefont {Birnkammer}},
  \bibinfo {author} {\bibfnamefont {M.~K.}\ \bibnamefont {Joshi}}, \bibinfo
  {author} {\bibfnamefont {A.}~\bibnamefont {Bastianello}}, \bibinfo {author}
  {\bibfnamefont {R.}~\bibnamefont {Blatt}}, \bibinfo {author} {\bibfnamefont
  {M.}~\bibnamefont {Knap}},\ and\ \bibinfo {author} {\bibfnamefont {C.~F.}\
  \bibnamefont {Roos}},\ }\href@noop {} {\bibinfo {title} {Observation of
  magnon bound states in the long-range, anisotropic {H}eisenberg model}}
  (\bibinfo {year} {2022}),\ \Eprint {https://arxiv.org/abs/2212.03899}
  {arXiv:2212.03899 [quant-ph]} \BibitemShut {NoStop}%
\bibitem [{\citenamefont {Greschner}\ \emph {et~al.}(2018)\citenamefont
  {Greschner}, \citenamefont {Cardarelli},\ and\ \citenamefont
  {Santos}}]{2018_Greschner}%
  \BibitemOpen
  \bibfield  {author} {\bibinfo {author} {\bibfnamefont {S.}~\bibnamefont
  {Greschner}}, \bibinfo {author} {\bibfnamefont {L.}~\bibnamefont
  {Cardarelli}},\ and\ \bibinfo {author} {\bibfnamefont {L.}~\bibnamefont
  {Santos}},\ }\href {https://doi.org/10.1103/PhysRevA.97.053605} {\bibfield
  {journal} {\bibinfo  {journal} {Phys. Rev. A}\ }\textbf {\bibinfo {volume}
  {97}},\ \bibinfo {pages} {053605} (\bibinfo {year} {2018})}\BibitemShut
  {NoStop}%
\bibitem [{\citenamefont {Liu}\ \emph {et~al.}(2018)\citenamefont {Liu},
  \citenamefont {Garrison}, \citenamefont {Deng}, \citenamefont {Gong},\ and\
  \citenamefont {Gorshkov}}]{Liu_2018}%
  \BibitemOpen
  \bibfield  {author} {\bibinfo {author} {\bibfnamefont {F.}~\bibnamefont
  {Liu}}, \bibinfo {author} {\bibfnamefont {J.~R.}\ \bibnamefont {Garrison}},
  \bibinfo {author} {\bibfnamefont {D.-L.}\ \bibnamefont {Deng}}, \bibinfo
  {author} {\bibfnamefont {Z.-X.}\ \bibnamefont {Gong}},\ and\ \bibinfo
  {author} {\bibfnamefont {A.~V.}\ \bibnamefont {Gorshkov}},\ }\href
  {https://doi.org/10.1103/PhysRevLett.121.250404} {\bibfield  {journal}
  {\bibinfo  {journal} {Phys. Rev. Lett.}\ }\textbf {\bibinfo {volume} {121}},\
  \bibinfo {pages} {250404} (\bibinfo {year} {2018})}\BibitemShut {NoStop}%
\bibitem [{\citenamefont {Schneider}\ \emph {et~al.}(2012)\citenamefont
  {Schneider}, \citenamefont {Hackermüller}, \citenamefont {Ronzheimer},
  \citenamefont {Will}, \citenamefont {Braun}, \citenamefont {Best},
  \citenamefont {Bloch}, \citenamefont {Demler}, \citenamefont {Mandt},
  \citenamefont {Rasch},\ and\ \citenamefont {Rosch}}]{Schneider_2012}%
  \BibitemOpen
  \bibfield  {author} {\bibinfo {author} {\bibfnamefont {U.}~\bibnamefont
  {Schneider}}, \bibinfo {author} {\bibfnamefont {L.}~\bibnamefont
  {Hackermüller}}, \bibinfo {author} {\bibfnamefont {J.~P.}\ \bibnamefont
  {Ronzheimer}}, \bibinfo {author} {\bibfnamefont {S.}~\bibnamefont {Will}},
  \bibinfo {author} {\bibfnamefont {S.}~\bibnamefont {Braun}}, \bibinfo
  {author} {\bibfnamefont {T.}~\bibnamefont {Best}}, \bibinfo {author}
  {\bibfnamefont {I.}~\bibnamefont {Bloch}}, \bibinfo {author} {\bibfnamefont
  {E.}~\bibnamefont {Demler}}, \bibinfo {author} {\bibfnamefont
  {S.}~\bibnamefont {Mandt}}, \bibinfo {author} {\bibfnamefont
  {D.}~\bibnamefont {Rasch}},\ and\ \bibinfo {author} {\bibfnamefont
  {A.}~\bibnamefont {Rosch}},\ }\href {https://doi.org/10.1038/nphys2205}
  {\bibfield  {journal} {\bibinfo  {journal} {Nat. Phys.}\ }\textbf {\bibinfo
  {volume} {8}},\ \bibinfo {pages} {213} (\bibinfo {year} {2012})}\BibitemShut
  {NoStop}%
\bibitem [{\citenamefont {Ronzheimer}\ \emph {et~al.}(2013)\citenamefont
  {Ronzheimer}, \citenamefont {Schreiber}, \citenamefont {Braun}, \citenamefont
  {Hodgman}, \citenamefont {Langer}, \citenamefont {McCulloch}, \citenamefont
  {Heidrich-Meisner}, \citenamefont {Bloch},\ and\ \citenamefont
  {Schneider}}]{Ronzheimer_2013}%
  \BibitemOpen
  \bibfield  {author} {\bibinfo {author} {\bibfnamefont {J.~P.}\ \bibnamefont
  {Ronzheimer}}, \bibinfo {author} {\bibfnamefont {M.}~\bibnamefont
  {Schreiber}}, \bibinfo {author} {\bibfnamefont {S.}~\bibnamefont {Braun}},
  \bibinfo {author} {\bibfnamefont {S.~S.}\ \bibnamefont {Hodgman}}, \bibinfo
  {author} {\bibfnamefont {S.}~\bibnamefont {Langer}}, \bibinfo {author}
  {\bibfnamefont {I.~P.}\ \bibnamefont {McCulloch}}, \bibinfo {author}
  {\bibfnamefont {F.}~\bibnamefont {Heidrich-Meisner}}, \bibinfo {author}
  {\bibfnamefont {I.}~\bibnamefont {Bloch}},\ and\ \bibinfo {author}
  {\bibfnamefont {U.}~\bibnamefont {Schneider}},\ }\href
  {https://doi.org/10.1103/PhysRevLett.110.205301} {\bibfield  {journal}
  {\bibinfo  {journal} {Phys. Rev. Lett.}\ }\textbf {\bibinfo {volume} {110}},\
  \bibinfo {pages} {205301} (\bibinfo {year} {2013})}\BibitemShut {NoStop}%
\bibitem [{\citenamefont {Yu}\ \emph {et~al.}(2017)\citenamefont {Yu},
  \citenamefont {Sun},\ and\ \citenamefont {Zhai}}]{Yu_2017}%
  \BibitemOpen
  \bibfield  {author} {\bibinfo {author} {\bibfnamefont {J.}~\bibnamefont
  {Yu}}, \bibinfo {author} {\bibfnamefont {N.}~\bibnamefont {Sun}},\ and\
  \bibinfo {author} {\bibfnamefont {H.}~\bibnamefont {Zhai}},\ }\href
  {https://doi.org/10.1103/PhysRevLett.119.225302} {\bibfield  {journal}
  {\bibinfo  {journal} {Phys. Rev. Lett.}\ }\textbf {\bibinfo {volume} {119}},\
  \bibinfo {pages} {225302} (\bibinfo {year} {2017})}\BibitemShut {NoStop}%
\bibitem [{\citenamefont {Viebahn}\ \emph {et~al.}(2021)\citenamefont
  {Viebahn}, \citenamefont {Minguzzi}, \citenamefont {Sandholzer},
  \citenamefont {Walter}, \citenamefont {Sajnani}, \citenamefont {G\"org},\
  and\ \citenamefont {Esslinger}}]{Viebahn_2021}%
  \BibitemOpen
  \bibfield  {author} {\bibinfo {author} {\bibfnamefont {K.}~\bibnamefont
  {Viebahn}}, \bibinfo {author} {\bibfnamefont {J.}~\bibnamefont {Minguzzi}},
  \bibinfo {author} {\bibfnamefont {K.}~\bibnamefont {Sandholzer}}, \bibinfo
  {author} {\bibfnamefont {A.-S.}\ \bibnamefont {Walter}}, \bibinfo {author}
  {\bibfnamefont {M.}~\bibnamefont {Sajnani}}, \bibinfo {author} {\bibfnamefont
  {F.}~\bibnamefont {G\"org}},\ and\ \bibinfo {author} {\bibfnamefont
  {T.}~\bibnamefont {Esslinger}},\ }\href
  {https://doi.org/10.1103/PhysRevX.11.011057} {\bibfield  {journal} {\bibinfo
  {journal} {Phys. Rev. X}\ }\textbf {\bibinfo {volume} {11}},\ \bibinfo
  {pages} {011057} (\bibinfo {year} {2021})}\BibitemShut {NoStop}%
\bibitem [{\citenamefont {Eckardt}(2017)}]{Eckardt_2017}%
  \BibitemOpen
  \bibfield  {author} {\bibinfo {author} {\bibfnamefont {A.}~\bibnamefont
  {Eckardt}},\ }\href {https://doi.org/10.1103/RevModPhys.89.011004} {\bibfield
   {journal} {\bibinfo  {journal} {Rev. Mod. Phys.}\ }\textbf {\bibinfo
  {volume} {89}},\ \bibinfo {pages} {011004} (\bibinfo {year}
  {2017})}\BibitemShut {NoStop}%
\bibitem [{\citenamefont {Islam}\ \emph {et~al.}(2015)\citenamefont {Islam},
  \citenamefont {Ma}, \citenamefont {Preiss}, \citenamefont {Tai},
  \citenamefont {Lukin}, \citenamefont {Rispoli},\ and\ \citenamefont
  {Greiner}}]{Islam_2015}%
  \BibitemOpen
  \bibfield  {author} {\bibinfo {author} {\bibfnamefont {R.}~\bibnamefont
  {Islam}}, \bibinfo {author} {\bibfnamefont {R.}~\bibnamefont {Ma}}, \bibinfo
  {author} {\bibfnamefont {P.~M.}\ \bibnamefont {Preiss}}, \bibinfo {author}
  {\bibfnamefont {M.~E.}\ \bibnamefont {Tai}}, \bibinfo {author} {\bibfnamefont
  {A.}~\bibnamefont {Lukin}}, \bibinfo {author} {\bibfnamefont
  {M.}~\bibnamefont {Rispoli}},\ and\ \bibinfo {author} {\bibfnamefont
  {M.}~\bibnamefont {Greiner}},\ }\href {https://doi.org/10.1038/nature15750}
  {\bibfield  {journal} {\bibinfo  {journal} {Nature}\ }\textbf {\bibinfo
  {volume} {528}},\ \bibinfo {pages} {77} (\bibinfo {year} {2015})}\BibitemShut
  {NoStop}%
\bibitem [{\citenamefont {Alicea}\ \emph {et~al.}(2011)\citenamefont {Alicea},
  \citenamefont {Oreg}, \citenamefont {Refael}, \citenamefont {von Oppen},\
  and\ \citenamefont {Fisher}}]{Alicea_2011}%
  \BibitemOpen
  \bibfield  {author} {\bibinfo {author} {\bibfnamefont {J.}~\bibnamefont
  {Alicea}}, \bibinfo {author} {\bibfnamefont {Y.}~\bibnamefont {Oreg}},
  \bibinfo {author} {\bibfnamefont {G.}~\bibnamefont {Refael}}, \bibinfo
  {author} {\bibfnamefont {F.}~\bibnamefont {von Oppen}},\ and\ \bibinfo
  {author} {\bibfnamefont {M.~P.~A.}\ \bibnamefont {Fisher}},\ }\href
  {https://doi.org/10.1038/nphys1915} {\bibfield  {journal} {\bibinfo
  {journal} {Nat. Phys.}\ }\textbf {\bibinfo {volume} {7}},\ \bibinfo {pages}
  {412} (\bibinfo {year} {2011})}\BibitemShut {NoStop}%
\bibitem [{\citenamefont {Sterdyniak}\ \emph {et~al.}(2012)\citenamefont
  {Sterdyniak}, \citenamefont {Regnault},\ and\ \citenamefont
  {M\"oller}}]{Sterdyniak_2012}%
  \BibitemOpen
  \bibfield  {author} {\bibinfo {author} {\bibfnamefont {A.}~\bibnamefont
  {Sterdyniak}}, \bibinfo {author} {\bibfnamefont {N.}~\bibnamefont
  {Regnault}},\ and\ \bibinfo {author} {\bibfnamefont {G.}~\bibnamefont
  {M\"oller}},\ }\href {https://doi.org/10.1103/PhysRevB.86.165314} {\bibfield
  {journal} {\bibinfo  {journal} {Phys. Rev. B}\ }\textbf {\bibinfo {volume}
  {86}},\ \bibinfo {pages} {165314} (\bibinfo {year} {2012})}\BibitemShut
  {NoStop}%
\bibitem [{\citenamefont {Palm}\ \emph {et~al.}(2021)\citenamefont {Palm},
  \citenamefont {Buser}, \citenamefont {L\'eonard}, \citenamefont
  {Aidelsburger}, \citenamefont {Schollw\"ock},\ and\ \citenamefont
  {Grusdt}}]{Palm_2021}%
  \BibitemOpen
  \bibfield  {author} {\bibinfo {author} {\bibfnamefont {F.~A.}\ \bibnamefont
  {Palm}}, \bibinfo {author} {\bibfnamefont {M.}~\bibnamefont {Buser}},
  \bibinfo {author} {\bibfnamefont {J.}~\bibnamefont {L\'eonard}}, \bibinfo
  {author} {\bibfnamefont {M.}~\bibnamefont {Aidelsburger}}, \bibinfo {author}
  {\bibfnamefont {U.}~\bibnamefont {Schollw\"ock}},\ and\ \bibinfo {author}
  {\bibfnamefont {F.}~\bibnamefont {Grusdt}},\ }\href
  {https://doi.org/10.1103/PhysRevB.103.L161101} {\bibfield  {journal}
  {\bibinfo  {journal} {Phys. Rev. B}\ }\textbf {\bibinfo {volume} {103}},\
  \bibinfo {pages} {L161101} (\bibinfo {year} {2021})}\BibitemShut {NoStop}%
\bibitem [{\citenamefont {Léonard}\ \emph {et~al.}(2022)\citenamefont
  {Léonard}, \citenamefont {Kim}, \citenamefont {Kwan}, \citenamefont
  {Segura}, \citenamefont {Grusdt}, \citenamefont {Repellin}, \citenamefont
  {Goldman},\ and\ \citenamefont {Greiner}}]{leonard2022realization}%
  \BibitemOpen
  \bibfield  {author} {\bibinfo {author} {\bibfnamefont {J.}~\bibnamefont
  {Léonard}}, \bibinfo {author} {\bibfnamefont {S.}~\bibnamefont {Kim}},
  \bibinfo {author} {\bibfnamefont {J.}~\bibnamefont {Kwan}}, \bibinfo {author}
  {\bibfnamefont {P.}~\bibnamefont {Segura}}, \bibinfo {author} {\bibfnamefont
  {F.}~\bibnamefont {Grusdt}}, \bibinfo {author} {\bibfnamefont
  {C.}~\bibnamefont {Repellin}}, \bibinfo {author} {\bibfnamefont
  {N.}~\bibnamefont {Goldman}},\ and\ \bibinfo {author} {\bibfnamefont
  {M.}~\bibnamefont {Greiner}},\ }\href@noop {} {\bibinfo {title} {Realization
  of a fractional quantum {H}all state with ultracold atoms}} (\bibinfo {year}
  {2022}),\ \Eprint {https://arxiv.org/abs/2210.10919} {arXiv:2210.10919
  [cond-mat.quant-gas]} \BibitemShut {NoStop}%
\bibitem [{\citenamefont {Goldman}\ and\ \citenamefont
  {Dalibard}(2014)}]{Goldman_2014}%
  \BibitemOpen
  \bibfield  {author} {\bibinfo {author} {\bibfnamefont {N.}~\bibnamefont
  {Goldman}}\ and\ \bibinfo {author} {\bibfnamefont {J.}~\bibnamefont
  {Dalibard}},\ }\href {https://doi.org/10.1103/PhysRevX.4.031027} {\bibfield
  {journal} {\bibinfo  {journal} {Phys. Rev. X}\ }\textbf {\bibinfo {volume}
  {4}},\ \bibinfo {pages} {031027} (\bibinfo {year} {2014})}\BibitemShut
  {NoStop}%
\bibitem [{\citenamefont {Bukov}\ \emph {et~al.}(2015)\citenamefont {Bukov},
  \citenamefont {D{\textquotesingle}Alessio},\ and\ \citenamefont
  {Polkovnikov}}]{Bukov_2015}%
  \BibitemOpen
  \bibfield  {author} {\bibinfo {author} {\bibfnamefont {M.}~\bibnamefont
  {Bukov}}, \bibinfo {author} {\bibfnamefont {L.}~\bibnamefont
  {D{\textquotesingle}Alessio}},\ and\ \bibinfo {author} {\bibfnamefont
  {A.}~\bibnamefont {Polkovnikov}},\ }\href
  {https://doi.org/10.1080/00018732.2015.1055918} {\bibfield  {journal}
  {\bibinfo  {journal} {Adv. Phys.}\ }\textbf {\bibinfo {volume} {64}},\
  \bibinfo {pages} {139} (\bibinfo {year} {2015})}\BibitemShut {NoStop}%
\bibitem [{\citenamefont {Cardarelli}(2019)}]{cardarelli_quantum_2019}%
  \BibitemOpen
  \bibfield  {author} {\bibinfo {author} {\bibfnamefont {L.}~\bibnamefont
  {Cardarelli}},\ }\href {https://doi.org/10.15488/9225} {\bibinfo {type}
  {{thesis}}},\ \bibinfo  {school} {Leibniz University Hannover} (\bibinfo
  {year} {2019})\BibitemShut {NoStop}%
\bibitem [{\citenamefont {Ma}\ \emph {et~al.}(2011)\citenamefont {Ma},
  \citenamefont {Tai}, \citenamefont {Preiss}, \citenamefont {Bakr},
  \citenamefont {Simon},\ and\ \citenamefont {Greiner}}]{Ma_2011}%
  \BibitemOpen
  \bibfield  {author} {\bibinfo {author} {\bibfnamefont {R.}~\bibnamefont
  {Ma}}, \bibinfo {author} {\bibfnamefont {M.~E.}\ \bibnamefont {Tai}},
  \bibinfo {author} {\bibfnamefont {P.~M.}\ \bibnamefont {Preiss}}, \bibinfo
  {author} {\bibfnamefont {W.~S.}\ \bibnamefont {Bakr}}, \bibinfo {author}
  {\bibfnamefont {J.}~\bibnamefont {Simon}},\ and\ \bibinfo {author}
  {\bibfnamefont {M.}~\bibnamefont {Greiner}},\ }\href
  {https://doi.org/10.1103/PhysRevLett.107.095301} {\bibfield  {journal}
  {\bibinfo  {journal} {Phys. Rev. Lett.}\ }\textbf {\bibinfo {volume} {107}},\
  \bibinfo {pages} {095301} (\bibinfo {year} {2011})}\BibitemShut {NoStop}%
\bibitem [{\citenamefont {Hartmann}\ \emph {et~al.}(2004)\citenamefont
  {Hartmann}, \citenamefont {Keck}, \citenamefont {Korsch},\ and\ \citenamefont
  {Mossmann}}]{Hartmann_2004}%
  \BibitemOpen
  \bibfield  {author} {\bibinfo {author} {\bibfnamefont {T.}~\bibnamefont
  {Hartmann}}, \bibinfo {author} {\bibfnamefont {F.}~\bibnamefont {Keck}},
  \bibinfo {author} {\bibfnamefont {H.~J.}\ \bibnamefont {Korsch}},\ and\
  \bibinfo {author} {\bibfnamefont {S.}~\bibnamefont {Mossmann}},\ }\href
  {https://doi.org/10.1088/1367-2630/6/1/002} {\bibfield  {journal} {\bibinfo
  {journal} {New J. Phys.}\ }\textbf {\bibinfo {volume} {6}},\ \bibinfo {pages}
  {2} (\bibinfo {year} {2004})}\BibitemShut {NoStop}%
\bibitem [{\citenamefont {Sträter}\ and\ \citenamefont
  {Eckardt}(2016)}]{Str_ter_2016}%
  \BibitemOpen
  \bibfield  {author} {\bibinfo {author} {\bibfnamefont {C.}~\bibnamefont
  {Sträter}}\ and\ \bibinfo {author} {\bibfnamefont {A.}~\bibnamefont
  {Eckardt}},\ }\href {https://doi.org/10.1515/zna-2016-0129} {\bibfield
  {journal} {\bibinfo  {journal} {Z. Naturforsch. A}\ }\textbf {\bibinfo
  {volume} {71}},\ \bibinfo {pages} {909} (\bibinfo {year} {2016})}\BibitemShut
  {NoStop}%
\bibitem [{\citenamefont {Sun}\ and\ \citenamefont {Eckardt}(2020)}]{Sun_2020}%
  \BibitemOpen
  \bibfield  {author} {\bibinfo {author} {\bibfnamefont {G.}~\bibnamefont
  {Sun}}\ and\ \bibinfo {author} {\bibfnamefont {A.}~\bibnamefont {Eckardt}},\
  }\href {https://doi.org/10.1103/PhysRevResearch.2.013241} {\bibfield
  {journal} {\bibinfo  {journal} {Phys. Rev. Res.}\ }\textbf {\bibinfo {volume}
  {2}},\ \bibinfo {pages} {013241} (\bibinfo {year} {2020})}\BibitemShut
  {NoStop}%
\bibitem [{\citenamefont {Kolovsky}\ \emph {et~al.}(2002)\citenamefont
  {Kolovsky}, \citenamefont {Ponomarev},\ and\ \citenamefont
  {Korsch}}]{Kolovsky_2002}%
  \BibitemOpen
  \bibfield  {author} {\bibinfo {author} {\bibfnamefont {A.~R.}\ \bibnamefont
  {Kolovsky}}, \bibinfo {author} {\bibfnamefont {A.~V.}\ \bibnamefont
  {Ponomarev}},\ and\ \bibinfo {author} {\bibfnamefont {H.~J.}\ \bibnamefont
  {Korsch}},\ }\href {https://doi.org/10.1103/PhysRevA.66.053405} {\bibfield
  {journal} {\bibinfo  {journal} {Phys. Rev. A}\ }\textbf {\bibinfo {volume}
  {66}},\ \bibinfo {pages} {053405} (\bibinfo {year} {2002})}\BibitemShut
  {NoStop}%
\end{thebibliography}%
